\documentclass[aps,nofootinbib,showpacs,eqsecnum,floats,preprint]{revtex4}

\begin{document}
\title{A One-Parameter Family of Time-Symmetric Initial Data for the Radial
Infall of a Particle into a Schwarzschild Black Hole}
\author{Karl Martel\thanks{martelk@physics.uoguelph.ca} and Eric Poisson\thanks{poisson@physics.uoguelph.ca}}
\affiliation{Department of Physics, University of Guelph, Guelph,
         Ontario, Canada N1G 2W1}
\date{\today}

\begin{abstract}
A one-parameter family of time-symmetric initial data for the radial infall
of a particle into a Schwarzschild black hole is constructed within the
framework of black-hole perturbation theory.  The parameter 
measures the amount of gravitational radiation present on the initial spacelike 
surface.  These initial data sets are then evolved by integrating the 
Zerilli-Moncrief wave equation in the presence of the particle. Numerical
results for the gravitational waveforms and
their power spectra are presented;  we show that the choice of initial 
data strongly influences the waveforms, both in their shapes and 
their frequency content. We also calculate the total energy radiated 
by the particle-black-hole system, as a function of the 
initial separation between the particle and the black hole, and as a function 
of the choice of initial data.  Our results confirm that for large initial 
separations, a conformally-flat initial three-geometry 
minimizes the initial gravitational-wave content, so that the total energy
radiated is also minimized.  For small 
initial separations, however, we show that the conformally-flat solution no longer 
minimizes the energy radiated.
\end{abstract}
\pacs{04.70.Bw, 04.30.Db, 04.30.Nk}
\maketitle

\section{Introduction}
Due to the advent of laser-interferometric gravitational-wave detectors,
there is a strong possibility that
gravitational waves from black-hole collisions will be detected in the
near future.  Black-hole mergers are good candidates for detection
because of the strength of the gravitational waves they emit, and of the
relatively high expected
event rate\cite{Zwart}. Detection of these collisions and successful extraction
of the black-hole parameters require a detailed theoretical understanding of the
gravitational waves emitted during the collision.
The lack of solutions to the dynamical two-body problem in general
relativity renewed interest in simulating black-hole collisions by numerically
integrating the full non-linear
Einstein field equations\cite{Cook_al,Gomez,Smarr,Alcubierre,Abrahams_al,Anninos1,Brandt_al,New,Baker1}.
The challenge is enormous.  For a recent review of the field of numerical relativity, see \cite{Lehner}.

From a numerical point of view, long-term simulations of black-hole collisions are limited by
the available memory\cite{Abrahams_al} and instabilities associated with the numerical implementation of
the full non-linear equations\cite{Gomez,Brandt_al,Miller}.
Because of these difficulties, numerical relativity is
not yet at the stage where it can simulate black-hole collisions for a
very long time\cite{New}.  Most simulations are started at a late stage of the
collision, when the two black holes are separated by a distance of just a few
black-hole masses.  These small initial separations imply that
non-trivial initial values must be provided for the
gravitational field in order to start the numerical evolution.

The initial-value problem consists of finding an initial three-metric $\gamma_{ij}$,
and an initial extrinsic curvature $K_{ij}$, that encode all of the physical information
about the system.  The solution should contain information about the state of
motion of the black holes, and information about the gravitational-wave content
of the initial three-surface. The initial gravitational-wave content is
important because it contains information about the motion of the black holes
prior to the beginning of the numerical evolution. (In the
past of the initial hypersurface, the two black holes emitted gravitational
waves that must be accounted for in the initial data.)  As we will see in
Sec.~\ref{sec:III}, the initial gravitational-wave content can play an
important role in the modeling of the gravitational waveforms.

For the case of colliding black holes, various methods have been developed to find initial data
that satisfy the Hamiltonian and momentum constraints of general relativity, e.g.,
the apparent-horizon method\cite{Thornburg}, conformal-imaging method\cite{Cook_Choptuik},
and puncture method\cite{Brandt_B}.

The starting point in these methods is a maximally embedded initial spacelike slice ($K^{i}_{\,\, i}=0$).  On this slice, the geometry
is described by  a conformally-flat
initial metric, $\gamma_{ij}=\Psi^{4}\bar{\gamma}_{ij}$, and a rescaled traceless extrinsic curvature
$K_{ij}= \Psi^{-2} \bar{K}_{ij}$, where $\bar{\gamma}_{ij}$ is the metric of three-dimensional flat space, $\Psi$ the
conformal factor, and $\bar{K}_{ij}$ the traceless conformal extrinsic curvature.
With these choices, the momentum constraints in vacuum can be recast in terms of quantities defined in the flat space: $\bar{\nabla}^{j}\bar{K}_{ij}=0$,
where $\bar{\nabla}^{i}$ is the flat-space covariant derivative.  Bowen and York obtained solutions to
this equation\cite{Bowen} in terms of a vector field $\bar{V}^{i}$ that contains information about
the spins and linear momenta of the holes on the initial surface, i.e.
$\bar{K}^{ij}=\bar{\nabla}^{i}\bar{V}^{j}+\bar{\nabla}^{j}\bar{V}^{i}-2/3\bar{\gamma}^{ij}\bar{\nabla}^{k}\bar{V}_{k}$;
explicit forms for $\bar{V}^{i}$ can be found in \cite{Cook}.
The three techniques mentioned above rely on the Bowen-York solutions to construct $\bar{K}_{ij}$, but because of the
different choices of topology for the initial hypersurface, the $\bar{K}_{ij}$'s they obtain are different.  But since the
differences in initial topology are mostly hidden behind the event horizons of the black holes,
they cannot affect significantly the resulting dynamics.  To reflect this,
we will generically refer to a $K_{ij}$ based on the Bowen-York solutions
as a ``longitudinal extrinsic curvature''.

Thus, solutions to the initial-value problem can be generated,
but the real problem is to construct physically suitable solutions.  This is a difficult problem because of the
non-linearities inherent to the theory.  It is very hard in practice to identify
which components of $\gamma_{ij}$ and $K_{ij}$ are to be constrained, and which are to be associated with dynamical
and gauge degrees of freedom.  It is now widely accepted that choosing a conformally-flat $\gamma_{ij}$ and
a longitudinal $K_{ij}$ are unlikely to yield physically relevant initial data sets.

For example, the assumption that the initial metric of a black-hole binary is conformally flat is likely to be inappropriate,
because the metric is known not to be conformally flat at the second post-Newtonian order\cite{Cook}.
As well, any physical black hole is likely to be rotating and, in this case,
the assumption of conformal flatness yields poor initial data.  The reason for this is well known:
Garat and Price\cite{Garat_P} have shown that there is no
spatial slicing of the Kerr solution that is both axisymmetric and conformally flat,
and reduces smoothly to the Schwarzschild solution in the no-rotation limit.
The techniques mentioned above can nevertheless provide solutions to the initial-value problem that
represent a rotating black hole, but the solutions cannot correspond to a stationary
Kerr black hole; there must be some gravitational radiation on the initial slice.

Even in the case of a perturbed Schwarzschild black hole, where conformally-flat slices can be constructed,
the solutions obtained with a longitudinal extrinsic curvature are not necessarily
adequate.  Lousto and Price\cite{LP2} have shown that for the head-on collision of two nonrotating
black holes in which one of the holes is much less massive than the other,
conformally-flat $\gamma_{ij}$ and longitudinal $K_{ij}$ data (CFL data) do not reproduce the numerical results\cite{LP2}.
In their analysis, they imposed the CFL data at a time $t_{o}$ and evolved it
forward in time.  They then looked
at the conditions at a later time $t_{1}$, and found that the extrinsic curvature
extracted from the numerical data matched poorly with the extrinsic curvature
obtained at $t=t_{1}$ from CFL data.
Instead, they found that the extrinsic curvature was
better represented by postulating a convective time
derivative\cite{LP3}, which means essentially that the time derivative of
the metric is proportional to the four-velocity of the small black hole.

The various problematic issues associated with the initial-value problem can be better understood
by having recourse to approximate methods\cite{Anninos,Khanna_al}.
For large initial separations and slow-motion processes, post-Newtonian (PN)
theory is useful, especially when the PN metric of two point masses is matched, in a buffer region,
with the metric of two distorted Schwarzschild black holes\cite{Alvi}.  For small mass ratios,
black-hole perturbation theory can be very
useful\cite{Pullin,Baker,Abrahams,PriPul}, since in this case there is no restriction on
the velocity of the small body.

This paper is the first of a series in which, following Lousto and Price\cite{LP2,LP3,LP1},
the initial-value problem will be
systematically explored in black-hole perturbation theory.  Our goal is to explore
the space of solutions to better identify physically realistic situations
(within the domain of validity).  This, we hope, will help us understand the
essential features that should be possessed by a solution to the initial-value
problem.  In this paper we look at the
radial infall of a particle starting from rest at a radius $r_{o}$
outside a Schwarzschild black hole, and we construct a one-parameter family of
solutions to the initial-value problem.  The parameter, $\alpha$, measures
the amount of radiation present on the initial hypersurface.  For $\alpha=1$ the initial
metric is conformally flat, a choice of initial data that was previously considered
by Lousto and Price\cite{LP1}.  Starting from our family of initial data, we calculate
the time evolution of the gravitational perturbations produced by the infalling particle.
We then calculate the gravitational waveforms, their power spectra, and the total energy radiated.
In this way, we are able to explore the limitations of the assumption of conformal flatness, and
determine when the contamination of the gravitational waveforms, due to unphysical radiation on the
initial three-surface, becomes important.

The paper is organized as follows.  In Sec.~\ref{sec:II} we give a short
overview of black-hole perturbation theory, appropriately restricted to the specific case of
radial infall into a Schwarzschild black hole.   The evolution of the metric
perturbations, governed by an inhomogeneous wave equation, is described in
Sec.~\ref{sec:ZMth}.  Our one-parameter family of solutions to the initial-value
problem is constructed in Sec.~\ref{sec:inith}.  Section \ref{sec:III} is
divided into four subsections.  In Sec.~\ref{sec:nummet} we explain the
numerical method used to integrate the wave equation
in the presence of the particle.  In Sec.~\ref{sec:waves} we present
the gravitational waveforms extracted from the simulations;
we discuss the influence of the choice of initial data on the waveforms.
In Sec.~\ref{sec:spectrum} we calculate the power spectrum of the waveforms for
selected values of $\alpha$ and initial separation $r_{o}$.  In Sec.~\ref{sec:energy}
we compute the total energy radiated in the $l=2$, 3, and 4 multipole moments of
the gravitational radiation field.  In Sec.~\ref{sec:V} we summarize our
findings and offer some conclusions. Finally, the Appendix contains a discussion of
the convergence properties of our numerical code.

Throughout the paper we use MTW conventions\cite{MTW} for the metric signature and
curvature tensors.  Greek indices run from 0 to 3, and Latin indices
run over the $t$ and $r$ Schwarzschild coordinates.  Geometrized units are used: $G=c=1$.

\section{Perturbation Theory} \label{sec:II}
\subsection{Zerilli-Moncrief Wave Equation} \label{sec:ZMth}
We consider the radial infall of a particle of mass $\mu$, starting from
rest at a radius $r_{o}$, into a Schwarzschild black hole.
The particle creates a small perturbation
in the gravitational field of the black hole and the total metric is expressed
as $g_{\mu \nu}=g^{\mathrm{Sch}}_{\mu \nu}+h_{\mu \nu}$, where $g_{\mu\nu}^{\mathrm{Sch}}$ is the
Schwarzschild metric and $h_{\mu \nu}$ is the perturbation; this can be
decomposed into odd-parity (axial) and even-parity (polar) modes, each carrying
spherical-harmonic indices $l$ and $m$\cite{Zerilli}.  Because of the azimuthal symmetry of
the problem, only the polar perturbations are excited by the infalling particle.
Our calculations are performed in the Regge-Wheeler
gauge\cite{RW} in which the perturbation --- for a specific $l$-pole and with $m=0$ ---
is written as
\begin{eqnarray}
h_{\mu \nu}^{l0}d x^{\mu}d x^{\nu}&=&Y^{l0}(\theta)\Big[(1-2M/r)H^{l0}_{0}(r,t){\mathrm d}t^2
+2H^{l0}_{1}(r,t){\mathrm d}t{\mathrm d}r\nonumber\\
&+&(1-2M/r)^{-1}H^{l0}_{2}(r,t){\mathrm d}r^2
+r^2K^{l0}(r,t){\mathrm d}\Omega^{2}\Big],\label{eqn:metric}
\end{eqnarray}
where $H^{l0}_{0}$, $H^{l0}_{1}$, $H^{l0}_{2}$, and $K^{l0}$ are the multipole
moments of the metric perturbations.  Below we will omit the superscripts
``$l0$'' where there is no risk of confusion.

Inserting the perturbed metric into the Einstein field equations yields
a set of seven coupled equations for the four metric perturbations;  three of
these come from the Hamiltonian and momentum constraints and the other
four are evolution equations. (The three remaining field equations describe
the axial perturbations of the Schwarzschild black hole.)  The equations are decoupled when written in
terms of the  gauge-invariant Zerilli-Moncrief function\cite{Moncrief}:
\begin{eqnarray}\label{eqn:psi}
\psi(r,t)&=&\frac{r}{\lambda+1}\Bigg\{K(r,t)
+\frac{f}{\Lambda}\Big[H_{2}(r,t)-r\frac{\partial}{\partial r}K(r,t)\Big]\Bigg\},\label{eqn:ZMfunction}
\end{eqnarray}
where $\lambda=(l+2)(l-1)/2$, $f=1-2M/r$,
and $\Lambda=\lambda+3M/r$.
This normalization for $\psi$ differs from the
normalization adopted in \cite{Anninos,PriPul,Cunning}.  A relation between
various normalizations used in the literature can be found in Lousto and Price\cite{LP1}.  Our
normalization is chosen to agree with theirs.

The evolution of $\psi(r,t)$ is governed by a single inhomogeneous
wave equation,
\begin{eqnarray} \label{eqn:wave}
\left[-\frac{\partial^2}{\partial t^2}+\frac{\partial^2}{\partial r^{*\,2}}-V(r)\right]\psi(r,t)=S(r,t),
\end{eqnarray}
where $r^{*}=r+2M\ln(r/2M-1)$ is the tortoise radius,
\begin{equation}\label{eqn:Vzm}
V(r)=\frac{2f}{r^2\Lambda^2}\left[\lambda^2(\Lambda+1)+\frac{9M^2}{r^2}\left(\Lambda-\frac{2M}{r}\right)\right]
\end{equation}
is the Zerilli-Moncrief potential, and
\begin{eqnarray} \label{eqn:Szm}
S(r,t)&=&\frac{2}{l(l+1)\Lambda}\Bigg\{r^2f\Big[f^2\frac{\partial}{\partial r}Q^{tt}-\frac{\partial}{\partial r}Q^{rr}\Big]\nonumber \\
&-&\frac{f^2}{\Lambda r}\Big[\lambda (\lambda-1)r^2
+(4\lambda -9)Mr+15M^2\Big]Q^{tt}
+r\Big(\Lambda-f\Big)Q^{rr}\Bigg\},
\end{eqnarray}
is the source term appropriate for radial infall.
The $Q^{ab}$'s ($a,b= t,r$) are constructed from the particle's stress-energy tensor:
\begin{equation} \label{eqn:Qab}
Q^{ab}(r,t)=8\pi\int T^{ab}(r,t,\theta,\varphi)Y^{*}(\theta,\varphi){\mathrm d}\Omega,
\end{equation}
where $T^{\alpha \beta}=\mu\int u^{\alpha}u^{\beta}\delta^4(x^{\nu}-x^{\nu}_{p}(\tau))\left(-g\right)^{-1/2}{\mathrm d}\tau$, $\tau$ is proper time
along the particle's world line $x_{p}^{\nu}(\tau)$,
$u^{\nu}=[\tilde{E}/f,-(\tilde{E}^2-f)^{1/2},0,0]$ is the four-velocity of the
particle, and $\tilde{E}$ is the conserved energy per unit mass.  We assume,
without loss of generality,
that the motion proceeds along the negative $z$ direction, so that $\theta=0$ on the
world line.  Evaluating the integrals in Eq.~(\ref{eqn:Qab}) and substituting
the results into Eq.~(\ref{eqn:Szm}), we obtain\footnote{The source term given here agrees
with the source term given in \cite{LP2,LP1} if we substitute ``$(1-2M/r_{p}(t))\delta^{'}(r-r_{p}(t))-2M/r^2\delta(r-r_{p}(t))$''
for ``$(1-2M/r)\delta^{'}(r-r_{p}(t))$'' in our source term.}
\begin{eqnarray} \label{eqn:Szm_sp}
S(r,t)&=&16\pi\frac{\sqrt{(2l+1)/(4\pi)}}{l(l+1)}\frac{\mu}{\tilde{E}}\frac{f^3}{\Lambda}\Bigg\{\delta^{'}(r-r_{p}(t))
-\Bigg[\frac{(\lambda+1)r-3M}{r^2 f}-\frac{6M\tilde{E}^{2}}{\Lambda r^2 f}\Bigg]\delta(r-r_{p}(t))\Bigg\},
\end{eqnarray}
where $r_{p}(t)$ describes the radial component of the particle's motion, a `` $'$ '' denotes
a derivative with respect to $r$, and we have used
$Y^{l0}(0)=\sqrt{(2l+1)/(4\pi)}$, which is appropriate
for motion along the $z$-axis.  For the radial infall of a particle starting
from rest, $r_{p}(t)$ is given implicitly by
\begin{eqnarray}\label{eqn:geod}
t&=&-4M\tilde{E}\Bigg\{-\frac{1}{2}\frac{r_{o}}{2M}\sqrt{\frac{r_{p}}{2M}}\sqrt{1-\frac{r_{p}}{r_{o}}}+\sqrt{\frac{r_{o}}{2M}}\left(1+\frac{r_{o}}{4M}\right)\arcsin \sqrt{\frac{r_{p}}{r_{o}}}\nonumber \\
&&+\frac{1}{4}\frac{1}{\tilde{E}^2}\ln\left[\frac{1+\frac{r_{p}}{2M}-2\frac{r_{p}}{r_{o}}-2\tilde{E}\sqrt{\frac{r_{p}}{2M}}\sqrt{1-\frac{r_{p}}{r_{o}}}}{1+\frac{r_{p}}{2M}-2\frac{r_{p}}{r_{o}}+2\tilde{E}\sqrt{\frac{r_{p}}{2M}}\sqrt{1-\frac{r_{p}}{r_{o}}}}\right]-\frac{\pi}{2}\sqrt{\frac{r_{o}}{2M}}\left(1+\frac{r_{o}}{4M}\right)\Bigg\},
\end{eqnarray}
where $r_{o}\equiv r_{p}(t=0)$ is the initial position of the particle, and $\tilde{E}=\sqrt{1-2M/r_{o}}$.
In Sec.~\ref{sec:III} we numerically integrate Eq.~(\ref{eqn:wave}) with
the source term of Eq.~(\ref{eqn:Szm_sp}).  The integration of
Eq.~(\ref{eqn:wave}) requires the specification
of initial data for the Zerilli-Moncrief function: both $\psi(r,t=0)$ and
$\frac{\partial}{\partial t}\psi(r,t=0)$ must be known.  We shall examine this
issue in the next subsection.

The Zerilli-Moncrief function $\psi(r,t)$ is directly related to the
gravitational waves received at infinity\footnote{With polarization axes
oriented along the $\theta$ and $\varphi$ directions, we have that the ``plus''
polarization of the gravitational field is given by $h_{+}=\frac{1}{2}\sqrt{(l+2)(l+1)l(l-1)}\psi (u) \,_{-2}Y^{l0}(\theta,\varphi)/r$,
where $_{-2}Y^{l0}$ are spin-weighted spherical harmonics\cite{Goldberg}.  The ``cross'' polarization,
in this orientation, is zero.}, and knowledge of $\psi(r,t)$ allows us to
calculate the rate at which gravitational waves carry energy to infinity.  For a
specific $l-$pole, the luminosity (the gravitational-wave flux integrated over
a two-sphere) is given by
\begin{eqnarray}\label{eqn:power}
\frac{{\mathrm d}}{{\mathrm d} u}E_{l}&=&\frac{1}{64\pi}\frac{(l+2)!}{(l-2)!}\left(\frac{\partial}{\partial u}\psi\right)^{2},
\end{eqnarray}
where $u=t-r^{*}$ is retarded time.  The total energy radiated is obtained
by integration:
\begin{eqnarray}\label{eqn:Eint}
E_{l}=\frac{1}{64\pi}\frac{(l+2)!}{(l-2)!}\int_{-\infty}^{\infty}{\mathrm d}u \left(\frac{\partial}{\partial u} \psi\right)^{2}.
\end{eqnarray}

\subsection{Initial Data}\label{sec:inith}
The initial-value problem of general relativity is solved once the metric
$\gamma_{ij}$ and the extrinsic curvature $K_{ij}$ are specified on a
three-dimensional spacelike hypersurface.  These quantities are
not freely specifiable, because they must satisfy the Hamiltonian and
momentum constraints of general relativity.  Because the constraints are non-linear,
finding solutions to the initial-value problem is a non-trivial task, and there is no
unique way of doing this.  Another essential difficulty is to find
solutions that represent a physical situation.  In the case
of black-hole collisions, $\gamma_{ij}$ and $K_{ij}$ must represent two moving
black holes with some initial amount of gravitational waves, as was explained in the Introduction.

Two much-studied solutions to the initial-value problem are the Misner solution\cite{Misner}
(which is generalized by the conformal-imaging method to situations where the initial
hypersurface is not a moment of time-symmetry)
and the Brill-Lindquist solution\cite{BrillL} (which is generalized by the puncture method).  They
represent the spatial metric of two black holes momentarily at rest, and
about to undergo a head-on collision.  Because the initial hypersurface
is a moment of time symmetry, $K_{ij}=0$ and the initial-value problem reduces to
finding $\gamma_{ij}$ (or the conformal factor).  These solutions and their generalizations were used as
initial data for the study of the head-on collision of two equal-mass black holes in full numerical
relativity\cite{Smarr,Anninos1,Abrahams}.

For the problem studied in this paper, namely the radial infall of a small mass
$\mu$ into a Schwarzschild black hole, we will concentrate on Brill-Lindquist (BL) type
initial data. (Our initial data sets are not taken to be inversion symmetric with respect
to the throat of the black hole.)  The Brill-Lindquist solution
can be expanded in powers of $\mu/M$ to yield an approximate solution to the initial-value problem appropriate for
perturbation theory; in this case $H_{1}=0$ and
\begin{equation}\label{eqn:Misner}
H_{2}=K=2 \mu \frac{\sqrt{4 \pi/(2l+1)}}{(1+M/2\bar{r}_{o})(1+M/2\bar{r})}\frac{\bar{r}_{<}^{l}}{\bar{r}_{>}^{l+1}},
\end{equation}
where $\bar{r}=r(1+\sqrt{f})^2/4$ is the isotropic radius, $f=1-2M/r$, and $\bar{r}_{<}$
($\bar{r}_{>}$) is the smaller (greater) of $\bar{r}$ and $\bar{r}_{o}$.  This
solution was first examined by Lousto and Price\cite{LP1}.  Here we wish to
construct a more general solution to the initial-value problem in perturbation
theory.

Our initial situation is that of a small mass $\mu$ momentarily at rest at a
radius $r_{o}$ outside a Schwarzschild black hole; this is a moment of
time symmetry, which implies that $K_{ij}=0$, so that the momentum constraints
are automatically satisfied.  To complete the solution to the initial-value
problem, we write the Hamiltonian constraint $^{3}R=8\pi\rho$ ($^{3}R$ is the
Ricci scalar on the initial hypersurface, and $\rho$ is the matter density) in
terms of the metric perturbations in the Regge-Wheeler gauge (Eq.~(\ref{eqn:metric})):
\begin{eqnarray}\label{eqn:cnst}
-r^2f\frac{{\mathrm d}^2}{{\mathrm d} r^2}K-(3r-5M)\frac{{\mathrm d}}{{\mathrm d}r}K+rf\frac{{\mathrm d}}{{\mathrm d}r}H_{2}+(H_{2}-K)\nonumber \\
+\frac{l(l+1)}{2}(H_{2}+K)=r^{2}f Q^{tt}=8\pi\sqrt{\frac{2l+1}{4\pi}}\mu\tilde{E}\delta(r-r_{o}).
\end{eqnarray}
This is a single equation for two unknowns: the metric functions $K(r)$ and
$H_{2}(r)$.  One of the functions can be specified freely, and we do this by
postulating the linear relationship $H_{2}=\alpha K$, where $\alpha$ is a constant.  This is motivated by
the fact that the relation $H_{2}=K$ produces a metric that is conformally flat:
\begin{eqnarray}
{\mathrm d}s^2|_{t=0}&=&\Big(1+K_{l0}(r)Y_{l0}(\theta)\Big)\left(f^{-1}{\mathrm d}r^2+r^2{\mathrm d}\Omega^2\right) \nonumber \\
&=&\Big(1+K_{l0}(r)Y_{l0}(\theta)\Big)\left(1+\frac{M}{2\bar{r}}\right)^{4}\left({\mathrm d}\bar{r}^2+\bar{r}^2{\mathrm d}\Omega^2\right).
\end{eqnarray}
Making the choice $H_{2}=K$ turns Eq.~(\ref{eqn:cnst}) into a
hypergeometric equation for $K(r)$, and as we shall show below, its solution
is given by Eq.~(\ref{eqn:Misner}) above.  The more general relation
$H_{2}=\alpha K$ also turns Eq.~(\ref{eqn:cnst}) into a hypergeometric
equation, for which we will be able to find solutions that generalize Eq.~(\ref{eqn:Misner}).
Our solutions are parameterized by $\alpha$, and the procedure outlined
here produces a one-parameter family of initial-data sets.  Setting $\alpha=1$
reproduces the BL solution of Eq.~(\ref{eqn:Misner}).

Postulating the relation $H_{2}=\alpha K$ at the initial moment $t=0$,
Eq.~(\ref{eqn:cnst}) becomes ($z\equiv r/2M$)
\begin{eqnarray}\label{eqn:vac}
z(1-z)\frac{{\mathrm d}^2}{{\mathrm d} z^2}K+\Big(5/2-\alpha-(3-\alpha)z\Big)\frac{{\mathrm d}}{{\mathrm d} z}K -\left((1-\alpha)-\frac{l(l+1)}{2}(1+\alpha)\right)K=0,
\end{eqnarray}
on either side of the particle's initial position ($r\neq r_{o}$).  This is the
hypergeometric equation.
For $r<r_{o}$ we must choose a solution that is regular at $r=2M$, but that
is allowed to diverge at $r=\infty$.  We denote this solution by $K_{<}(r)$.
For $r>r_{o}$ we must choose a solution that is regular at $r=\infty$, but
that may diverge at $r=2M$.  This solution is denoted by $K_{>}(r)$.  In terms
of these, the solution for $K(r)$ takes the form\footnote{In principle, we are free to add any
multiple of $K_{>}(r)$ to our particular solution. This represents a different choice of initial
three-geometry\cite{LP2}, and we do not consider this possibility here.}
\begin{eqnarray} \label{eqn:Kt}
K(r,0)=C(r_{o})K_{<}(r_{<})K_{>}(r_{>}),
\end{eqnarray}
where $r_{<}$ ($r_{>}$)
is the smaller (greater) of $r$ and $r_{o}$,
\begin{eqnarray}\label{eqn:K<>}
K_{<}(r)&=&\left(\frac{r}{2M}\right)^{-b}F\left(b,b-c+1;1-a+b;\frac{2M}{r}\right), \nonumber \\
K_{>}(r)&=&\left(\frac{r}{2M}\right)^{-a}F\left(a,a-c+1;1-b+a;\frac{2M}{r}\right), \label{eqn:K>}
\end{eqnarray}
$F(,;;)$ is the hypergeometric function,
\begin{eqnarray}\label{eqn:hyparg}
a&=&1-\frac{\alpha}{2}+\frac{1}{2}\Bigg[\alpha^2+2(1+\alpha)l(l+1)\Bigg]^{1/2}, \nonumber \\
b&=&1-\frac{\alpha}{2}-\frac{1}{2}\Bigg[\alpha^2+2(1+\alpha)l(l+1)\Bigg]^{1/2}, \nonumber \\
c&=&\frac{5}{2}-\alpha,
\end{eqnarray}
and $C(r_{o})$ is a constant ensuring
that $\frac{\partial}{\partial r}K$ has the correct discontinuity at $r=r_{o}$.
Integrating across the $\delta$-function in Eq.~(\ref{eqn:cnst}), using $H_{2}=\alpha K$ and
Eq.~(\ref{eqn:Kt}) for $K$, $C(r_{o})$ is easily found to be
\begin{eqnarray}\label{eqn:Const}
C(r_{o})&=&8\pi\frac{\mu}{\tilde{E}}\sqrt{\frac{2l+1}{4\pi}}\frac{1}{r_{o}^2}\left[K_{<}(r_{o})K^{'}_{>}(r_{o})-K^{'}_{<}(r_{o})K_{>}(r_{o})\right]^{-1} \nonumber \\
&=& 4\pi\sqrt{\frac{2l+1}{4\pi}}\frac{\mu}{M}\left(\frac{r_{o}}{2M}\right)^{1-\alpha}\frac{\Gamma(r_{o})}{\tilde{E}},
\end{eqnarray}
where
\begin{eqnarray} \label{eqn:DelF}
\Gamma(r_{o})&=&\Bigg[aF(a+1,a-c+1;1-b+a;2M/r_{o})F(b,b-c+1;1-a+b;2M/r_{o})\nonumber \\
&-&bF(a,a-c+1;1-b+a;2M/r_{o})F(b+1,b-c+1;1-a+b;2M/r_{o})\Bigg]^{-1}.
\end{eqnarray}
The second equality in Eq.~(\ref{eqn:Const}) follows from Eq.~(15.2.2) of
\cite{Stegun} together with Eq.~(\ref{eqn:K>}) to evaluate $K^{'}_{<}(r_{o})$
and $K^{'}_{>}(r_{o})$.

The initial value of the Zerilli-Moncrief function can now be obtained by
inserting $K(r)$ and $H_{2}=\alpha K(r)$ in Eq.~(\ref{eqn:ZMfunction}).  This,
together with $\frac{\partial}{\partial t}\psi(r,0)=0$, completely specifies
our family of initial-data sets for the radial infall of a particle starting
from rest.  In Sec.~\ref{sec:III} we numerically integrate the
Zerilli-Moncrief equation for this family of initial data.

The conformally-flat solution can easily be recovered as a special case of the
initial-data sets presented above.  For a conformally-flat initial
three-geometry ($\alpha=1$),
the parameters appearing in the hypergeometric functions (Eq.~(\ref{eqn:hyparg}))
take the values
$a=l+1$, $b=-l$, and $c=\frac{3}{2}$,
and Eq.~(\ref{eqn:K<>}) becomes
\begin{eqnarray}\label{eqn:K<>fl}
K_{<}(r)&=&\left(\frac{r}{2M}\right)^{l}F(-l,-l-1/2;-2l;2M/r) \nonumber \\
&=&2^{-(2l+1)}\left(\frac{r}{2M}\right)^{l}\left(1+\sqrt{f}\right)^{2l+1} \nonumber\\
&=&\left(\frac{\bar{r}}{2M}\right)^{l}\frac{1}{1+M/2\bar{r}},\\
K_{>}(r)&=& \left(\frac{r}{2M}\right)^{-(l+1)}F(l+1,l+1/2;2(l+1);2M/r) \nonumber \\
&=&2^{2l+1} \left(\frac{r}{2M}\right)^{-(l+1)}\left(1+\sqrt{f}\right)^{-(2l+1)} \nonumber \\
&=&\left(\frac{2M}{\bar{r}}\right)^{l+1}\frac{1}{1+M/2\bar{r}},
\end{eqnarray}
where we have used Eq.~(15.1.13) of \cite{Stegun}.  With these results, it is
a trivial matter to show that
\begin{eqnarray}\label{eqn:KK}
K_{<}(r_{<})K_{>}(r_{>})&=&\frac{2M}{(1+M/2\bar{r}_{o})(1+M/2\bar{r})}\frac{\bar{r}_{<}^{l}}{\bar{r}_{>}^{l+1}}.
\end{eqnarray}
With the conformally-flat values of $a$, $b$, and $c$, $\Gamma(r_{o})$, given
by  Eq.~(\ref{eqn:DelF}), becomes
\begin{eqnarray}
\Gamma(r_{o})&=& \Bigg[(l+1)F\left(l+2,l+1/2;2(l+1);2M/r_{o}\right)F\left(-l,-l-1/2;-2l;2M/r_{o}\right) \nonumber \\
&&+ lF\left(-l+1,-l-1/2;-2l;2M/r_{o}\right)F\left(l+1,l+1/2;2(l+1);2M/r_{o}\right)\Bigg]^{-1} \nonumber \\
&=&\frac{\tilde{E}}{2l+1}\Bigg[F(-l,-l-1/2;-2l;2M/r_{o})F(l+1,l+1/2;2(l+1);2M/r_{o})\Bigg]^{-1} \nonumber \\
&=& \frac{\tilde{E}}{2l+1},
\end{eqnarray}
where the second equality follows from the second identity of Eq.~(15.1.13) of
\cite{Stegun} applied to $F(l+2,;;)$ and $F(l+1,;;)$, and the third equality
from the first identity of Eq.~(15.1.13) of \cite{Stegun} applied to both
hypergeometric functions.  Substituting this results into Eq.~(\ref{eqn:Const})
yields
\begin{eqnarray}\label{eqn:Cfl}
C(r_{o})&=& \sqrt{\frac{4\pi}{2l+1}}\frac{\mu}{M}.
\end{eqnarray}
Finally, inserting Eq.~(\ref{eqn:KK}) and Eq.~(\ref{eqn:Cfl}) into Eq.~(\ref{eqn:Kt}) gives
\begin{eqnarray}
K(r)=2 \mu \frac{\sqrt{4 \pi/(2l+1)}}{(1+M/2\bar{r}_{o})(1+M/2\bar{r})}\frac{\bar{r}_{<}^{l}}{\bar{r}_{>}^{l+1}},
\end{eqnarray}
which is indeed the conformally-flat solution of Eq.~(\ref{eqn:Misner}).

In order for the perturbed metric to be real (as opposed to complex), the solutions to
Eq.~(\ref{eqn:vac}) must be real functions. Thus, $\alpha$ must be such that
the parameters $a$ and $b$, as listed in Eqs.~(\ref{eqn:hyparg}), are real.  It
is easy to verify that the quantity appearing under the square root will be
positive if $\alpha\geq \alpha_{+}$, or $\alpha\leq \alpha_{-}$, where
\begin{eqnarray}
\alpha_{\pm}=-l(l+1)\pm\sqrt{(l+2)(l+1)l(l-1)}.
\end{eqnarray}
It can be verified that $\alpha_{+}$ varies between $-1.10102$ (when $l=2$) and
$-1$ (for $l\rightarrow \infty$).  On the other hand, $\alpha_{-}$ monotically
decreases from $-10.8990$ (when $l=2$) as $l$ increases.  Hence, the metric
functions will be real for all values of $l$ if $\alpha$ is restricted by
$\alpha \geq -1$.

\begin{figure}[!t]
\vspace*{2.5in}
\special{hscale=31 vscale=31 hoffset=-13.0 voffset=190.0
         angle=-90.0 psfile=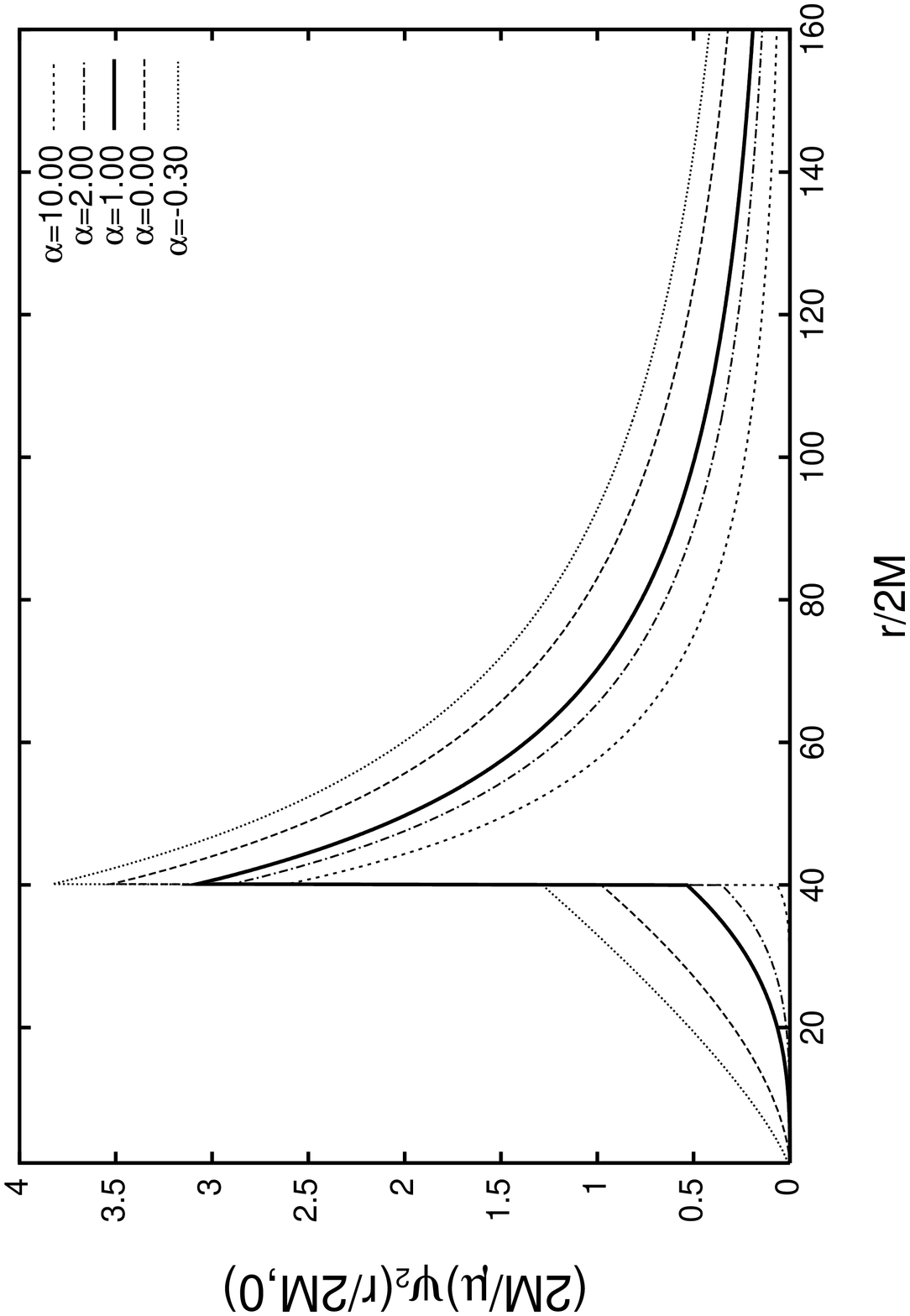}
\special{hscale=31 vscale=31 hoffset=225.0 voffset=190.0
         angle=-90.0 psfile=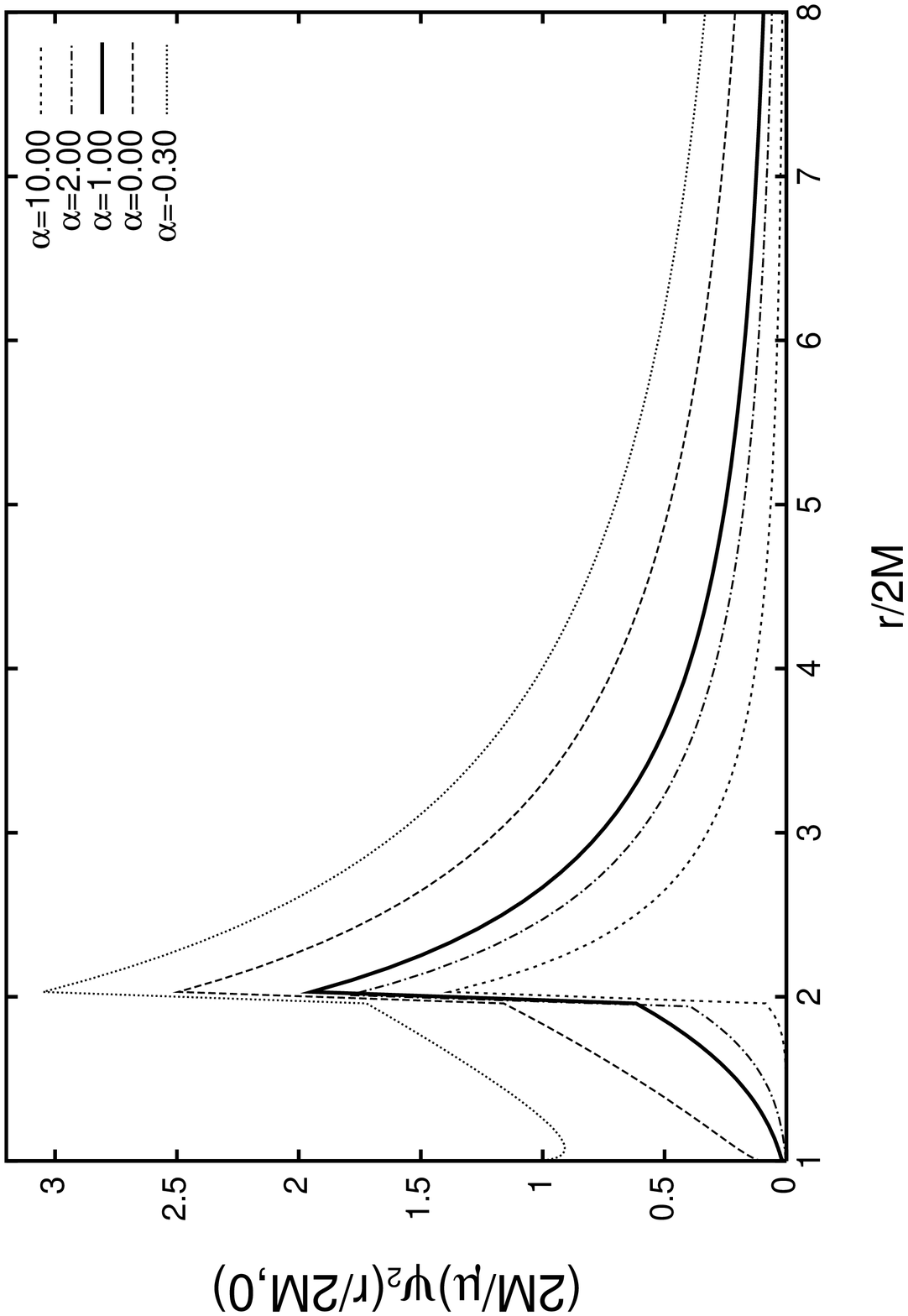}
\caption{Initial values of $\psi(r,t)$ for $l=2$; the particle is initially located at $r_{o}/2M=40$ (left) and $r_{o}/2M=2$ (right).  In both cases $\psi(r,0)$ is
peaked and discontinuous at the particle's location, $r=r_{o}$.
Decreasing (increasing) the value of $\alpha$ increases (decreases) the amplitude
and width of the initial pulse.  These properties can be associated with the amount
of gravitational radiation present on the initial hypersurface.  For fixed $\alpha$,
the amplitude of the peak decreases with decreasing $r_{o}$.}\label{fig:Initial}
\end{figure}

Figure~\ref{fig:Initial} displays the Zerilli-Moncrief function $\psi(r,0)$ for
the cases $r_{o}/2M=40$ and $r_{o}/2M=2$.  The figure shows
that $\psi(r,0)$ is peaked at the particle's location, and its
amplitude and width change with the value of $\alpha$;  the function is
discontinuous at $r=r_{o}$ because of the discontinuity in the term
$\partial K/\partial r$ appearing in Eq.~(\ref{eqn:psi}).  We see that the amplitude
of the peak increases with decreasing values of $\alpha$, and this effect is more
pronounced as $r_{o}$ decreases.  Intuitively, we associate a smaller
(larger) amplitude with a smaller (larger) amount of
gravitational radiation initially present in the spacetime.  Thus, to initial data
with a small (large) value of $\alpha$ we associate a large (small) amount of initial radiation.

\section{Numerical Results} \label{sec:III}
\subsection{Numerical Method} \label{sec:nummet}
We now describe the numerical method used to integrate Eq.~(\ref{eqn:wave}),
starting with the initial data constructed in the preceding section.  We recall
that the source term is given by Eq.~(\ref{eqn:Szm_sp}); special care must be
given to the fact that $S(r,t)$ is singular at $r=r_{p}(t)$, the trajectory of
the infalling particle.  A numerical method was presented by Lousto and
Price \cite{LP2} and we employ it here with a slight modification, which we discuss below.

The numerical domain is a staggered grid in $r^{*}$ and $t$, with a stepsize
$\Delta\equiv \Delta r^{*}/4M=\Delta t/2M$, over the region of spacetime
bounded by the spacelike hypersurface $t=0$ (the moment of time symmetry),
a null hypersurface approximating the event horizon [located at $u=T-r_{p}^{*}(T)$,
where $T$ is given by Eq.~(\ref{eqn:geod}) with $r_{p}(T)/2M =1.0001$]
\footnote{This choice is made to ensure that the particle's
contribution to the radiation is (almost) zero at the end of the numerical integration.
This means that the value of $u$ associated with the event horizon in our numerical
grid changes every time we change $r_{o}$.},
and a null hypersurface approximating future null-infinity [located
at $v/2M=(t+r^{*})/2M\approx 1500$].  The Zerilli-Moncrief function is extracted from the numerical data on this last
hypersurface, i.e. $\psi(r,t)$ is evaluated at $v/2M\approx 1500$ and expressed
as a function of $u=t-r^{*}$.  The results presented in Sec.~\ref{sec:waves} were obtained by setting $\Delta=0.01$ in
the evolution scheme presented below.

\begin{figure}[t]
\vspace*{2.5in}
\special{hscale=45 vscale=45 hoffset=-60.0 voffset=225.0
         angle=-90.0 psfile=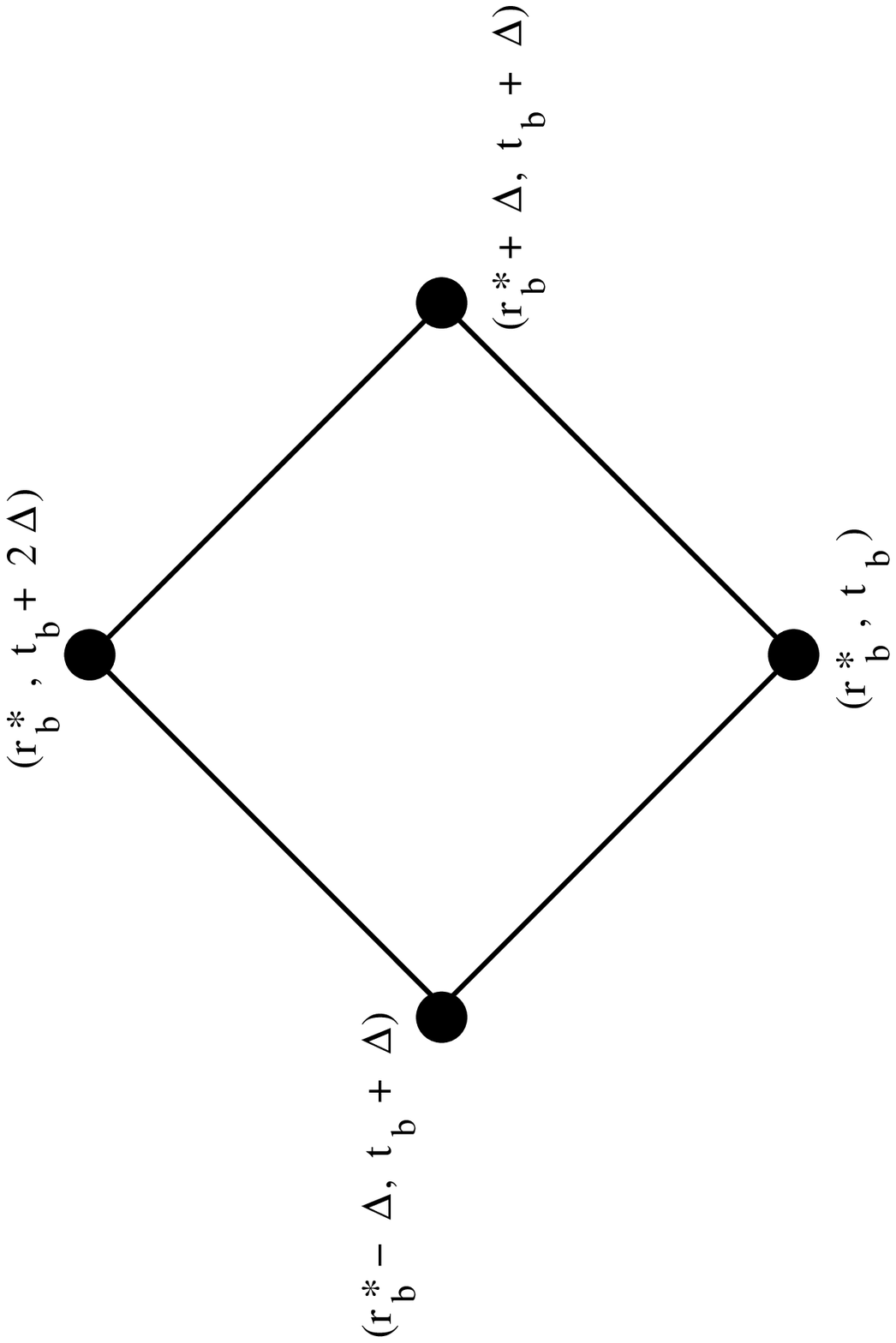}
\special{hscale=45 vscale=45 hoffset=170.0 voffset=225.0
         angle=-90.0 psfile=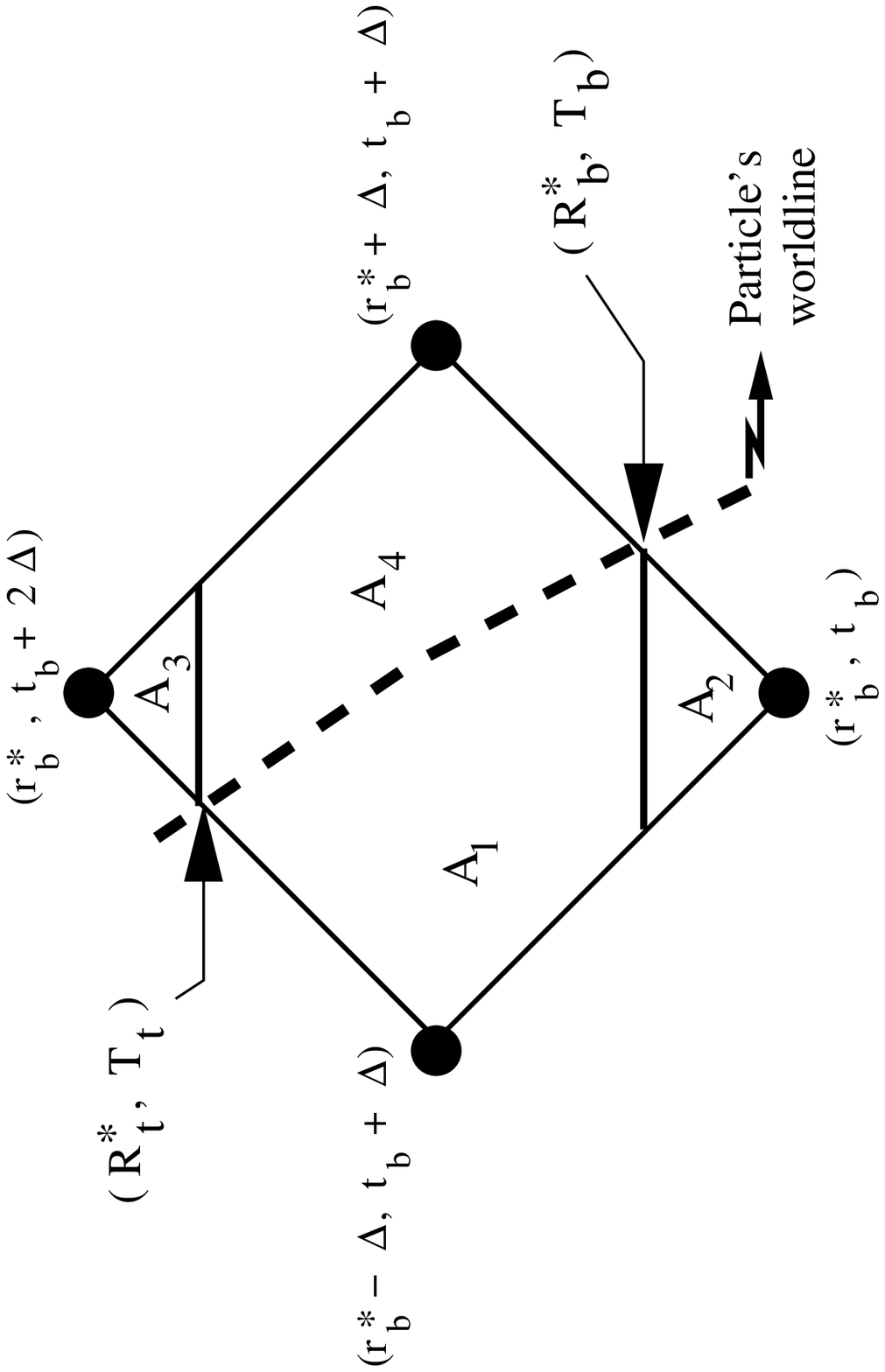}
\caption{The left cell is never crossed by the particle's trajectory.  In such
a cell the evolution of $\psi(r,t)$ is unaffected by the source term in
Eq.~(\ref{eqn:wave}), and Eq.~(\ref{eqn:fidif}) is used to evolve $\psi(t,r)$
forward in time.  The right cell is traversed by the particle's world line.
The areas shown in the cell are used in the numerical algorithm that incorporates
the effect of the source term in Eq.~(\ref{eqn:wave}).  The wave function is evolved
according to Eq.~(\ref{eqn:fidifS}).  The labels $(R^{*}_{b},T_{b})$ and
$(R^{*}_{t},T_{t})$ refer to the points ($r^{*},t$) at which the particle enters and
leaves the cell, respectively.} \label{fig:cell}
\end{figure}

The finite-difference method must take into account the source term, which is
non zero (and singular) on the world--line of the infalling particle.  The
numerical grid is divided into cells of area $4\Delta^{2}$. (Instead of using
the physical area $2\Delta^2$, we found it convenient, following Lousto and Price\cite{LP2},
to use $A=\int\mathrm{d}u\mathrm{d}v=4\Delta^2$.) The cells can be separated into two groups.  The first group
corresponds to cells in which $r\neq r_{p}(t)$ everywhere; these cells are
never traversed by the particle.  The second group is such that $r=r_{p}(t)$
somewhere within the cell; these cells are traversed by the infalling particle.
The two types of cells are displayed in
Fig.~\ref{fig:cell}; in the diagram, the particle enters the cell on the right of
$r^{*}_{b}$, and leaves on the left.

The evolution of $\psi(r^{*},t)$ across cells belonging to the first group
is not affected by the source term.  For these cells, we use a standard
scheme, accurate to $\mathcal{O}(\Delta^{4})$, for the homogeneous wave
equation in the presence of a potential:
\begin{eqnarray}\label{eqn:fidif}
\psi(r^{*}_{b},t_{b}+2\Delta)&=&-\psi(r^{*}_{b},t_{b})+\Big[\psi(r^{*}_{b}+\Delta,t_{b}+\Delta)+\psi(r^{*}_{b}-\Delta,t_{b}+\Delta)\Big]\left[1-\frac{\Delta}{2}^{2}V_{b}\right],
\end{eqnarray}
where $(r^{*}_{b},t_{b})$ designates the bottom corner of the cell
(see Fig.~\ref{fig:cell}), and $V_{b}=V(r^{*}_{b})$.

The evolution of $\psi(r,t)$ across cells
belonging to the second group is affected by the singular source term.  To obtain the evolution of $\psi$
across these cells, we closely follow Lousto and Price\cite{LP2}, carefully keeping terms up to order $\mathcal{O}(\Delta^{3})$.
Integrating Eq.~(\ref{eqn:wave}) over a grid cell, term by term, we get:
\begin{eqnarray}
\int \int \mathrm{d}A \left(-\frac{\partial^2}{\partial t^2}+\frac{\partial^2}{\partial r^{*2}}\right)\psi &=&-4\left[\psi(r^{*}_{b},t_{b}+2\Delta)+\psi(r^{*},t_{b})\right. \nonumber\\
&&\left.-\psi(r^{*}_{b}-\Delta,t_{b}+\Delta)-\psi(r^{*}_{b}+\Delta,t_{b}+\Delta)\right], \\
\int \int \mathrm{d}A \,V(r^{*})\psi&=&V(r^{*}_{b})\left(A_{1}\psi(r_{b}^{*}-\Delta,t_{b}+\Delta)+A_{2}\psi(r_{b}^{*},t_{b})\right. \nonumber \\
&+&\left.A_{3}\psi(r_{b}^{*},t_{b}+2\Delta)+A_{4}\psi(r_{b}^{*}+\Delta,t_{b}+\Delta)\right)+\mathcal{O}(\Delta^{3}),
\end{eqnarray}
where $\mathrm{d}A=\mathrm{d}u\,\mathrm{d}v$, and the $A_{i}$'s are areas dividing the cell into
four non-equal parts (as shown in Fig.~\ref{fig:cell}).  The integration of the source term over the cell
eliminates the $\delta$-function in Eq.~(\ref{eqn:Szm_sp})
and the term involving the derivative of the
$\delta$-function is evaluated by integrating by parts.  The result is
\begin{eqnarray}\label{eqn:Src_exp}
\int\int \mathrm{d}A\, S &=& -\kappa \int_{T_{b}}^{T_{t}} {\mathrm d} t \frac{f(t)}{\Lambda(t)^2}\left[\frac{6M}{r_{p}(t)}(1-\tilde{E}^2)+\lambda(\lambda+1)-\frac{3M^2}{r_{p}(t)^2}+4\lambda\frac{M}{r_{p}(t)}\right]\nonumber \\
&&\pm \kappa \Bigg\{\frac{f(T_{b})}{\Lambda(T_{b})}[1\mp \dot{r}^{*}_{p}(T_{b})]^{-1} + \frac{f(T_{t})}{\Lambda(T_{t})}[1\pm \dot{r}^{*}_{p}(T_{t})]^{-1}\Bigg\},
\end{eqnarray}
where $\kappa$ is defined by $l(l+1)\tilde{E}\kappa\equiv 16\pi\mu\sqrt{(2l+1)/(4\pi)}$,
$\dot{r}^{*}_{p}=-\sqrt{\tilde{E}^{2}-f}/\tilde{E}$, $f(t)\equiv f(r_{p}(t))$,
$\Lambda(t)\equiv\Lambda(r_{p}(t))$, $T_{b}$ is the time at
which the particle enters the cell, and $T_{t}$ the time at which it leaves the
cell.  In the previous expression, the upper (lower) sign for the first boundary
term (a function of $T_{b}$) applies when the particle enters the cell on the
right (left) of $r^{*}_{b}$.  Similarly, the upper (lower) sign for the second
boundary term (a function of $T_{t}$) applies when the particle leaves the cell
on the right (left) of $r^{*}_{b}$.

Substituting the previous results in Eq.~(\ref{eqn:wave}) and solving for $\psi(r_{b}^{*},t_{b}+2\Delta)$, we obtain
\begin{eqnarray}\label{eqn:fidifS}
\psi(r^{*}_{b},t_{b}+2\Delta)&=&-\psi(r^{*}_{b},t_{b})\Bigg[1+\frac{V_{b}}{4}(A_{2}-A_{3})\Bigg]\nonumber\\
&&+\psi(r^{*}_{b}+\Delta,t_{b}+\Delta)\Bigg[1-\frac{V_{b}}{4}(A_{1}+A_{3})\Bigg]\nonumber\\
&&+\psi(r^{*}_{b}-\Delta,t_{b}+\Delta)\Bigg[1-\frac{V{b}}{4}(A_{4}+A_{3})\Bigg]\nonumber\\
&&-\frac{1}{4}\left(1-\frac{V_{b}}{4}A_{3}\right)\int\int\mathrm{d}A S(r,t). \label{eqn:fds}
\end{eqnarray}
Notice that we have kept the factor of $\left(1-\frac{V_{b}}{4}A_{3}\right)$ in front of the source term.  This factor does not
appear in the scheme devised by Lousto and Price\cite{LP2}, but is required to obtain quadratic convergence:
the boundary terms appearing in Eq.~(\ref{eqn:Src_exp}) are $\mathcal{O}(1)$ in $\Delta$ and
$\frac{V_{b}}{4}A_{3}$ represents a correction of order $\mathcal{O}(\Delta^{2})$ to their contribution, which we cannot neglect if we want
quadratic convergence.

The previous equations cannot be used to evolve $\psi$ from the initial time $t=0$
to the next time $t=\Delta$.  To get $\psi(r^{*}_{b},\Delta)$ we need
$\psi(r^{*}_{b}-\Delta,0)$, $\psi(r^{*}_{b}+\Delta,0)$, and $\psi(r^{*}_{b},-\Delta)$.
By construction, $\psi(r^{*}_{b},-\Delta)$ is not known, but we can use the
simple relation $\psi(r^{*}_{b},-\Delta)=\psi(r^{*}_{b},\Delta)+\mathcal{O}(\Delta^{3})$
which holds by virtue of the fact that the data at $t=0$ is time-symmetric: $\partial \psi/\partial t=0$.
This relation is valid so long as $r^{*}_{p}(t)\neq r^{*}_{b}$ everywhere in the cell.
We chose $r^{*}_{b}-\Delta<r^{*}_{p}(0)<r^{*}_{b}$ which, for radial infall,
guarantees that the relation holds.

The gravitational waveform produced by the infalling particle is given directly
by $\psi(u)$, which is extracted from the numerical data in the way described
previously.  Other quantities of interest are the energy radiated and its power
spectrum.  We calculate these by first taking a fast Fourier transform of the
wave function $\psi(u)$.  This is given by
\begin{eqnarray}
\tilde{\psi}(\omega)&=&\int_{-\infty}^{\infty}{\mathrm d}\omega e^{i \omega u}\psi(u),
\end{eqnarray}
where $\omega$ is the frequency.  From Eq.~(\ref{eqn:power}) and Parceval's
theorem, we find that the power spectrum is
\begin{eqnarray}\label{eqn:spc}
\frac{{\mathrm d}}{{\mathrm d} \omega}E_{l}&=&\frac{1}{64\pi^2}\frac{(l+2)!}{(l-2)!}\omega^2\left|\tilde{\psi}(\omega)\right|^2.
\end{eqnarray}
The total energy radiated in each multipole moment is calculated by performing
a Romberg integration over all frequencies.

\subsection{Waveforms} \label{sec:waves}
\begin{figure}[!ht]
\vspace*{4.8in}
\special{hscale=31 vscale=31 hoffset=-13.0 voffset=358.0
         angle=-90.0 psfile=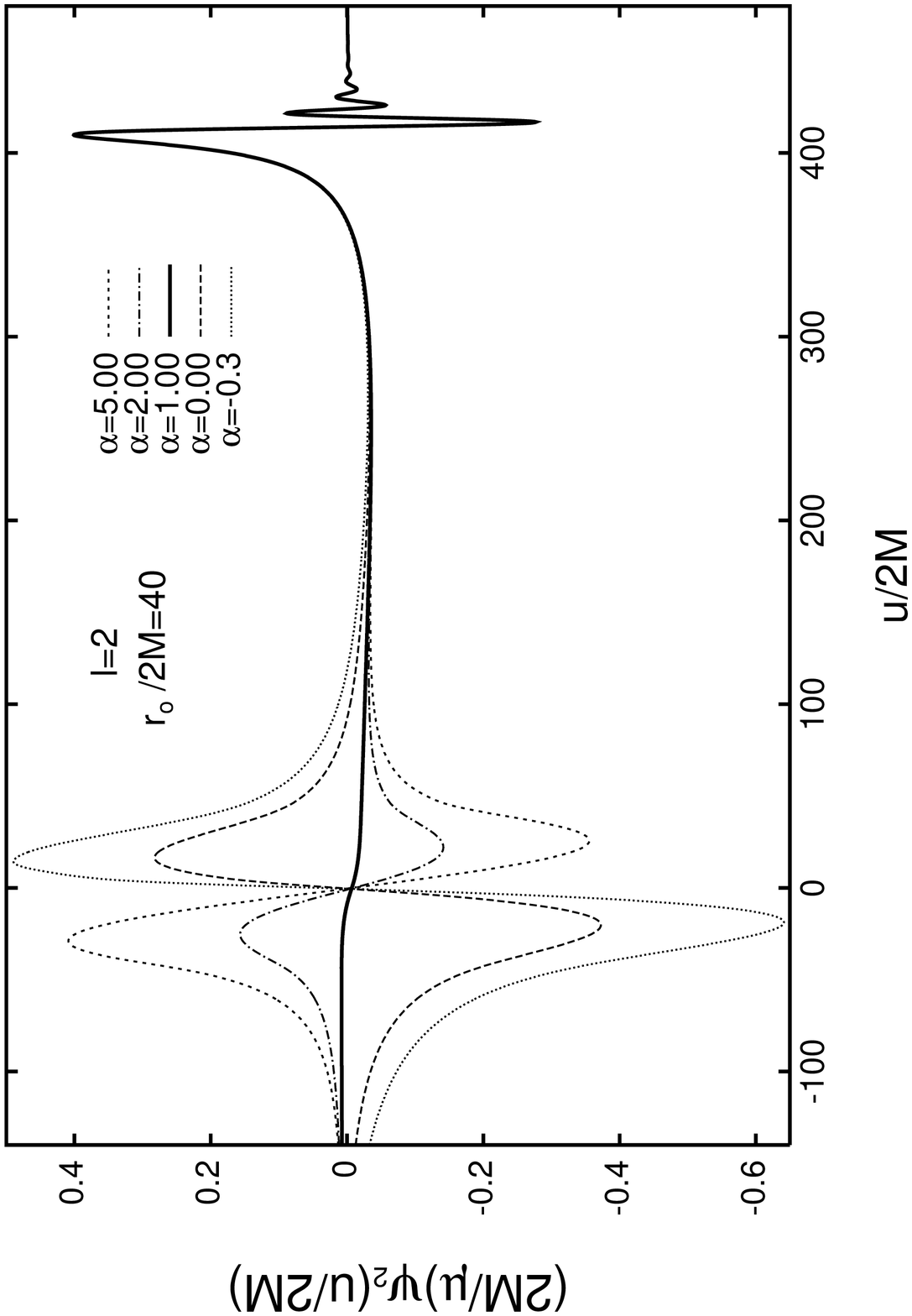}
\special{hscale=31 vscale=30 hoffset=225.0 voffset=358.0
         angle=-90.0 psfile=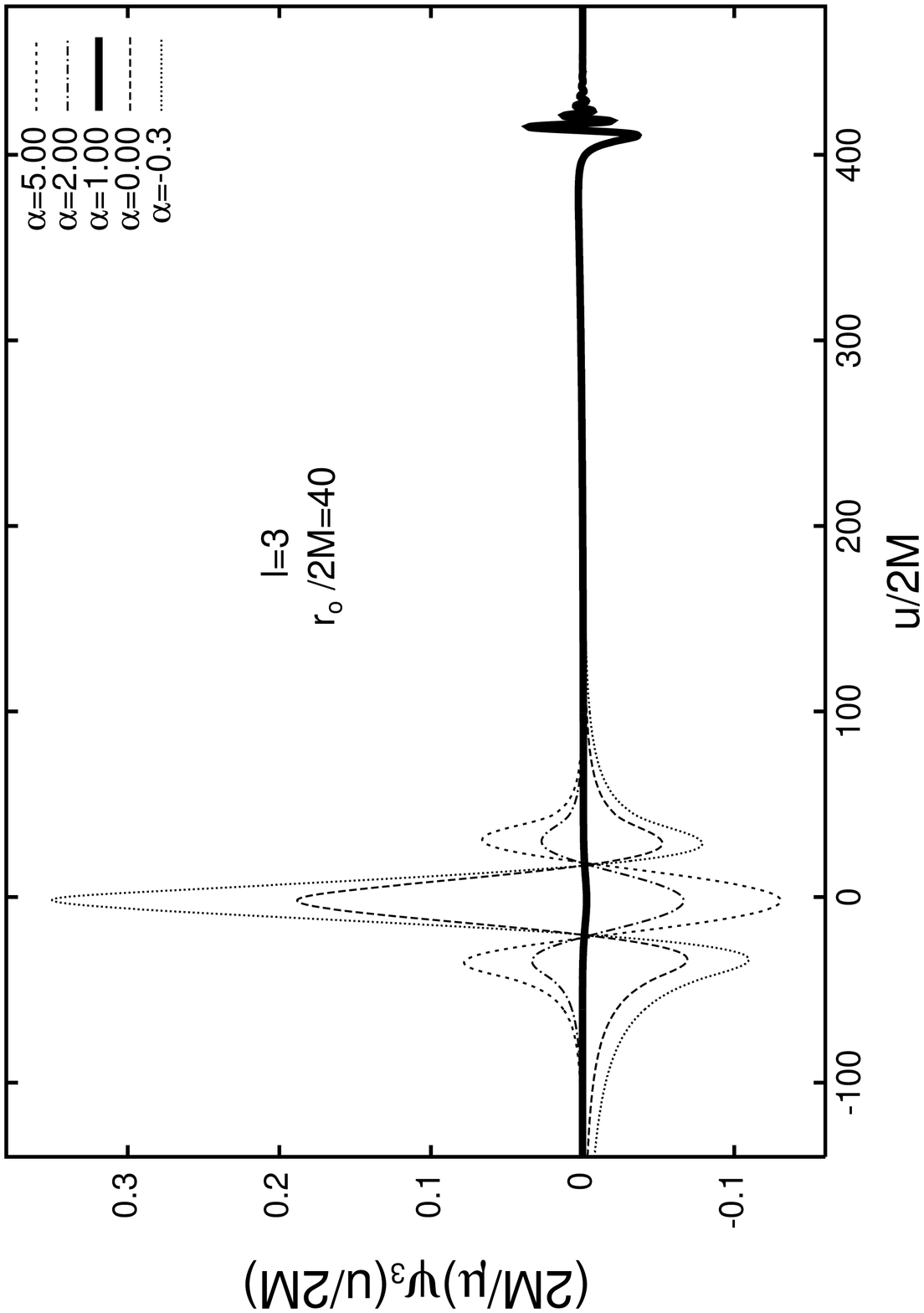}
\special{hscale=31 vscale=30 hoffset=-13.0 voffset=190.0
         angle=-90.0 psfile=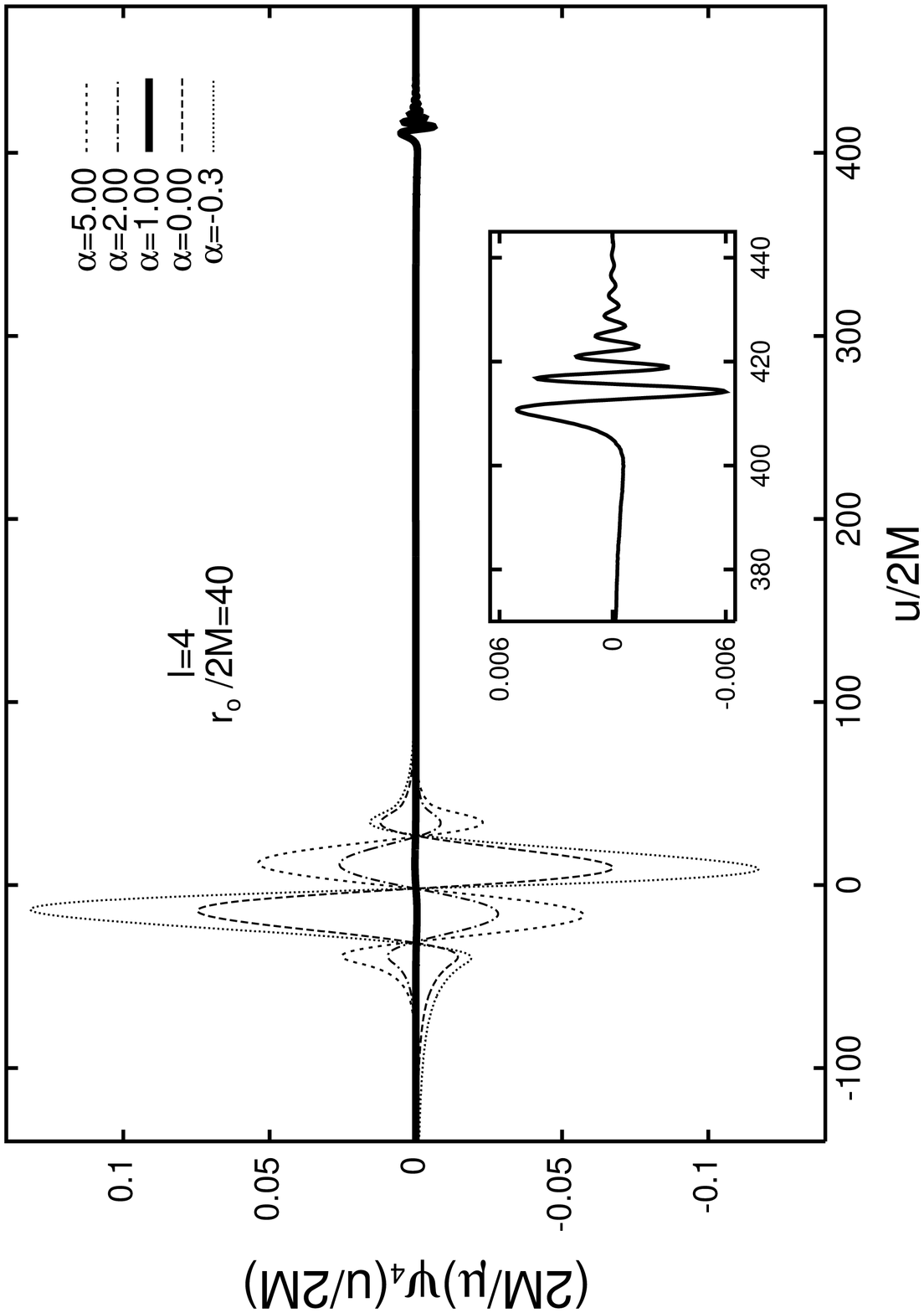}
\caption{The $l=2$, 3, and 4 modes of the Zerilli-Moncrief function for infall
from $r_{o}/2M=40$, for $\alpha= -0.3$, 0, 1, 2, and 5. For each multipole
moment, the early-time behaviour is
dominated by the initial data contribution ($u/2M<350$), while the late-time
behaviour is dominated by radiation emitted by the particle in the strong-field
region ($350<u/2M<400$) and by quasi-normal ringing of the black hole ($u/2M>400$).
The early-time portion of the waveforms (first stage) depends strongly on the parameter $\alpha$,
which labels the choice of initial data.  In contrast, the late-time portion (second and third stages)
are completely insensitive to the choice of initial data.} \label{fig:Pr40}
\end{figure}
The radiation arriving at the null boundary of our domain of integration contains radiation from
the initial data, radiation emitted by the particle as
it falls toward the black hole, and radiation corresponding to the black-hole's response
to the perturbation.  Because the particle is at rest at $t=0$, it
initially produces very little radiation.  The early part of the waveform is
therefore dominated by radiation contained in the initial data.  As time proceeds,
the radiation produced by the particle becomes noticeable, and the dynamics
associated with the perturbation of the event horizon starts to play a role.
The radiative process can therefore be separated into three stages.  The first stage is
associated with the initial data, the second with the infalling particle, and the
third with the event-horizon dynamics.  We now describe these stages in detail.

The initial data is time-symmetric, and it therefore consists of two radiation pulses:
one pulse is outgoing, and the other is incoming.  The outgoing pulse
proceeds to infinity, with some backscattering [which is small, unless the pulse
originates in the strong-field region of the spacetime ($r^{*}\leq 0$)].  On
the other hand, the ingoing pulse moves toward the black hole and is backscattered
by the potential barrier at $r^{*}\approx 0$.  The reflected pulse then proceeds
to infinity, where it arrives delayed with respect to the original outgoing pulse.
This is the first stage of the radiative process, and it is directly associated with
the initial data.  The second stage is associated
with the motion of the particle.  As the particle accelerates toward the
black hole, it produces radiation which propagates to infinity.  This
happens either by direct propagation or by backscattering from the potential
barrier.  The third and final stage of the radiative process is the response of the black
hole to the perturbation.  As the particle approaches
the black hole, it tidally deforms the event horizon, which becomes dynamical.  The
radiation produced in this process interacts strongly with the potential barrier
outside the black hole, and the result, after transmission to infinity, is a pattern
of damped oscillations.  This response of the black hole to the perturbation created by
the particle is known as quasi-normal ringing\cite{Leaver}.

As long as $r_{o}\gg 2M$, the different stages of the radiative process can be
clearly distinguished (initial-data pulses, acceleration radiation, and
quasi-normal ringing).  But when $r_{o}$ is comparable to $2M$, the three epochs become confused, and this gives rise to
interfering waveforms.  For this reason, varying the initial separation between the
particle and the black hole can have an important effect on the gravitational
waveforms.

Let us describe more fully the first stage of the radiative process.  For infall
from a large distance ($r_{o} \gg 2M$), the outgoing pulse travels
directly to future null-infinity with very little backscattering, since the pulse
originates in a region where the potential $V(r)$ is weak.  This is shown in the waveform
as a single pulse of radiation at early times
($u\approx - r^{*}_{o}$).  The ingoing pulse, on the other hand, proceeds toward the
black hole, where it is almost entirely backscattered by the potential barrier.
The reflected pulse then proceeds to infinity and this gives rise to a second pulse
of radiation at $u\approx r^{*}_{o}$.  Varying $\alpha$ changes the
amount of gravitational radiation initially present at $t=0$, and
the amplitude of the two pulses depends on $\alpha$.  This
can be seen in Fig.~\ref{fig:Pr40}, where the $l=2$, 3, and 4 modes of the
Zerilli-Moncrief function are displayed for infall from $r_{o}/2M=40$, for several values of $\alpha$.
For $l=2$, the two pulses have a minimum amplitude when
$\alpha=1$, and the amplitudes vary smoothly with $\alpha$.

At later times, once the pulses have made their way to
infinity, the radiative process becomes dominated by the particle's
contribution; this is the second stage.  At this time, the particle has
entered the strong-field region of the spacetime, the acceleration is large,
and it radiates strongly.  This stage lasts for a short time, because the
particle quickly falls into the black hole.  The burst of radiation from the
particle is then quickly replaced by the third stage, quasi-normal ringing.
When $r_{o}\gg 2M$, the acceleration radiation and the quasi-normal ringing
stages are insensitive to the choice of initial data.  This is illustrated in
Fig.~\ref{fig:Pr40}.

\begin{figure}[!ht]
\vspace*{4.8in}
\special{hscale=31  vscale=31 hoffset=-13.0 voffset=358.0
         angle=-90.0 psfile=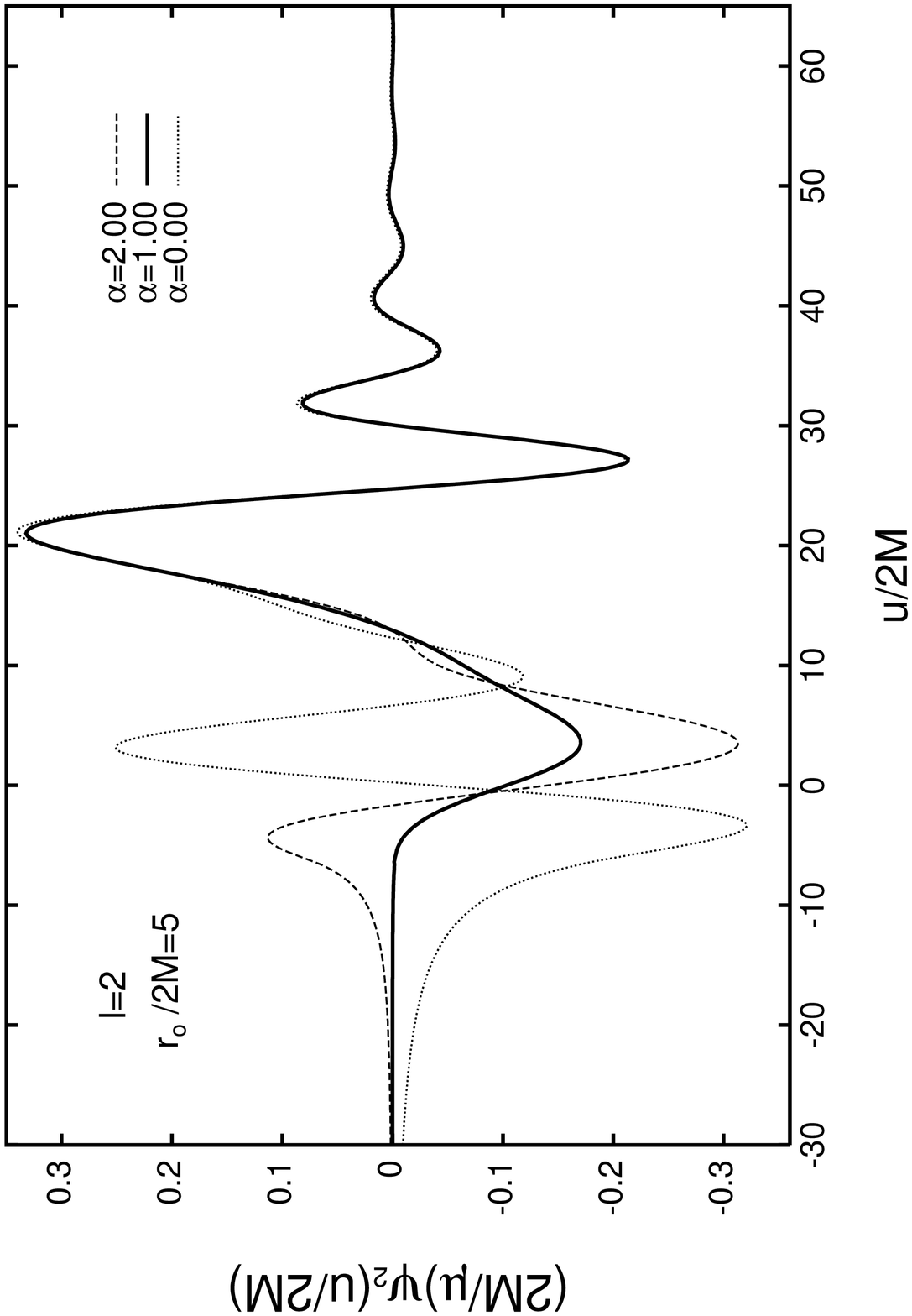}
\special{hscale=31 vscale=31 hoffset=225.0 voffset=358.0
         angle=-90.0 psfile=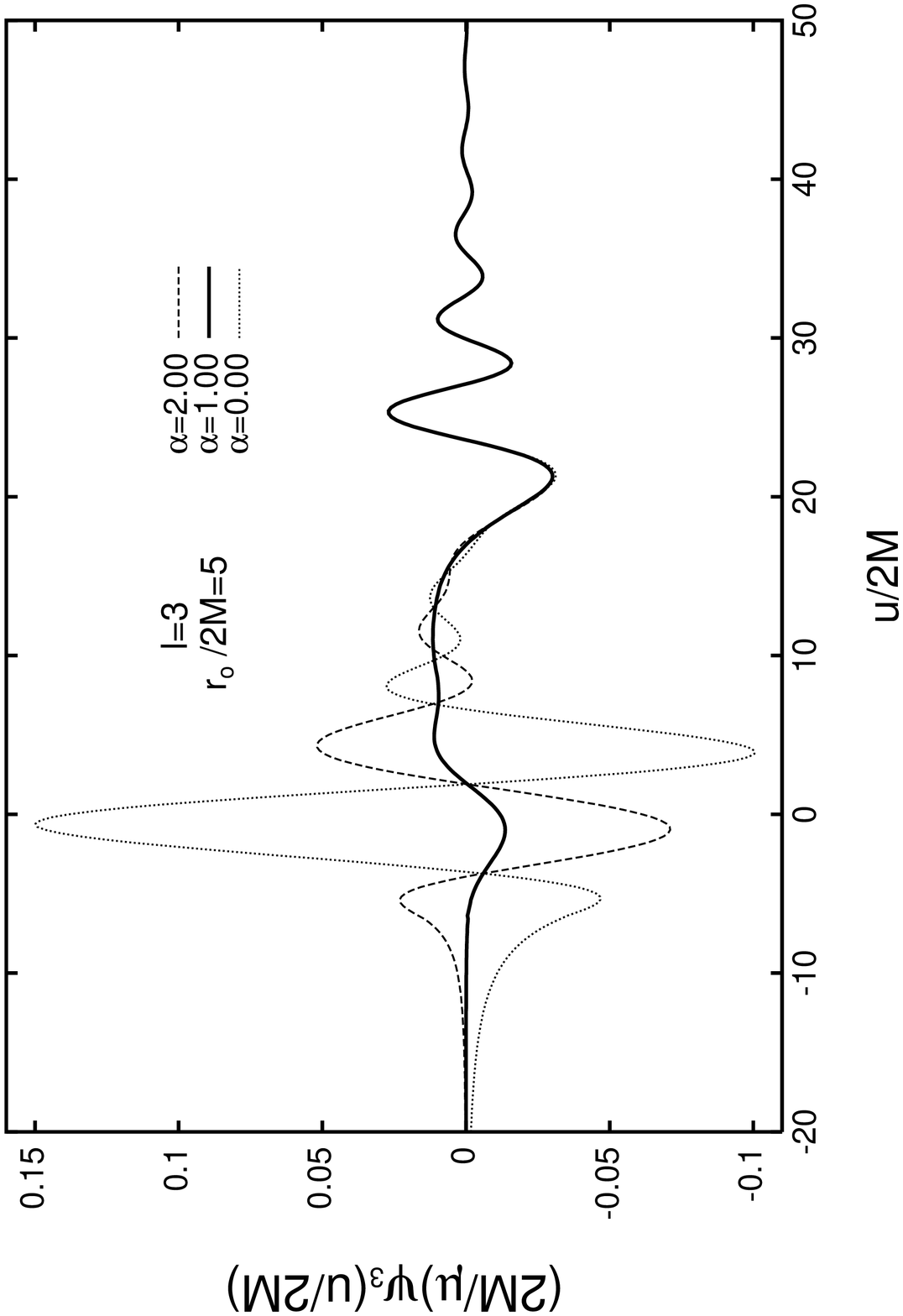}
\special{hscale=31 vscale=31 hoffset=-13.0 voffset=190.0
         angle=-90.0 psfile=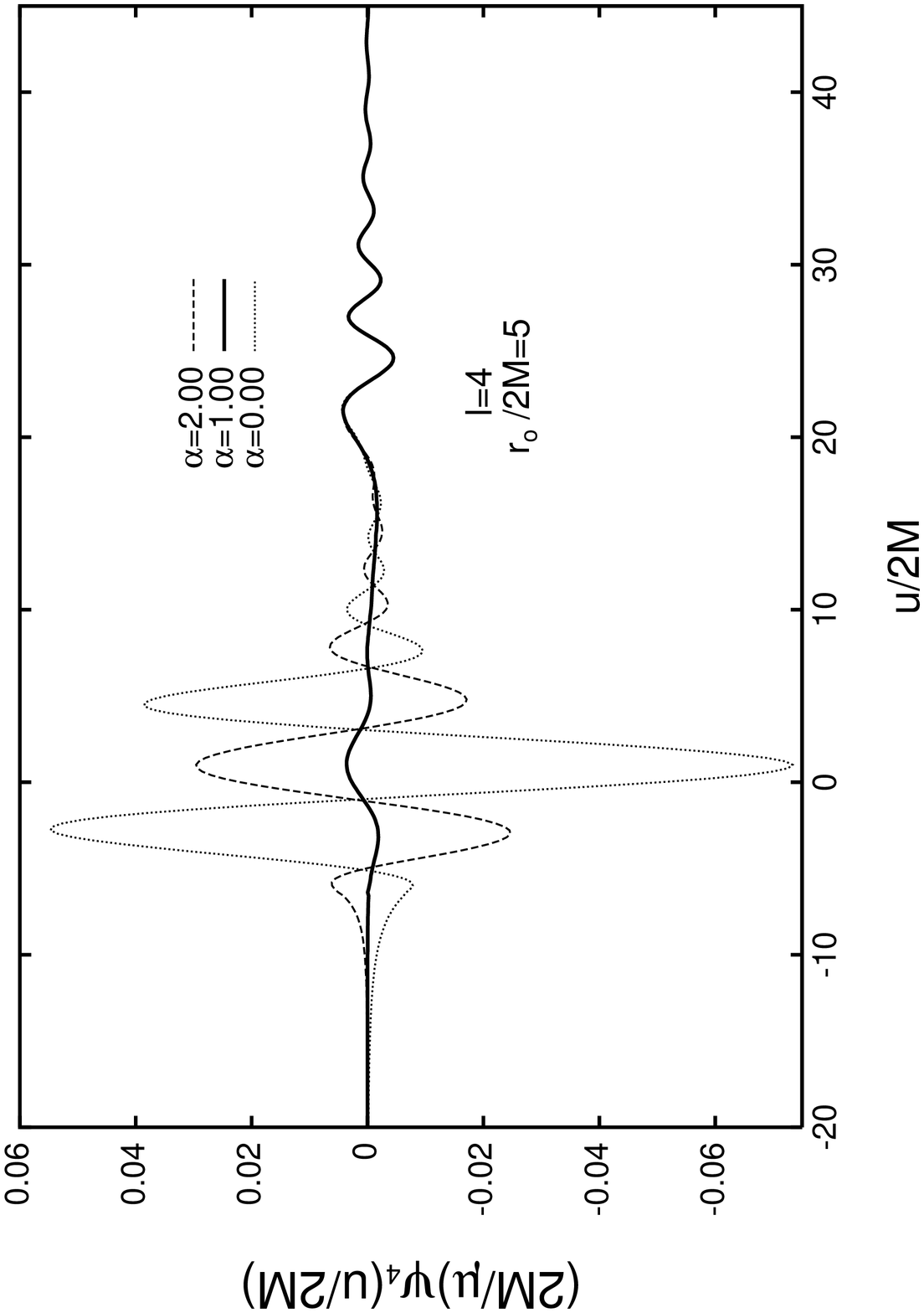}
\caption{The $l=2$, 3, and 4 modes of the Zerilli-Moncrief function for infall
from $r_{o}/2M=5$, for $\alpha=0$, 1, and 2.  The waveform changes smoothly as
$\alpha$ is varied away from 1 (conformal flatness), but the change is more dramatic
when $\alpha$ is decreased.  The three stages of the radiative process
now overlap and become confused.  However, we still witness an early-time sensitivity,
and a late-time insensitivity, to the choice of initial data.} \label{fig:Pr5}
\end{figure}
The situation changes when $r_{o}$ is chosen within the strong-field region
of the spacetime.  The outgoing pulse from the initial data still
proceeds directly to infinity with little backscattering, creating a pulse
at $u\approx -r_{o}^{*}$
(see Fig.~\ref{fig:Pr5}).  However, the ingoing pulse is no longer
entirely backscattered by the potential barrier;  part of the pulse is now
transmitted to the black hole.
The backscattered portion of the ingoing pulse proceeds to infinity where it
generates a second pulse at $u\approx r_{o}^{*}$, while the transmitted pulse
reaches the event horizon.  As a result, the event horizon becomes distorted, and starts
radiating into quasi-normal modes;
the second pulse is therefore immediately followed by an epoch of quasi-normal ringing excited by the
transmitted pulse.  The amplitude of the quasi-normal ringing depends
on the strength of the excitation, and is therefore highly sensitive to the choice
of initial data.  This early epoch of quasi-normal ringing cannot be seen
for infall from large $r_{o}$, because in that case the initial data contains
mostly low-frequency gravitational radiation that is almost totally reflected by the
Zerilli-Moncrief potential\cite{Chandra}.

For intermediate values of $r_{o}$ ($4\lesssim r_{o}/2M < 10$), the particle starts
in the strong-field region of the spacetime, and its large acceleration
causes it to radiate strongly almost immediately.  The second stage, which
for infall from large $r_{o}$ was dominated by acceleration radiation,
is now the sum of acceleration radiation and
quasi-normal ringing excited by the initial ingoing pulse, as was
discussed above.  This can be seen especially clearly for the $l=3$ and 4 modes
of the Zerilli-Moncrief function displayed in Fig.~\ref{fig:Pr5}, for infall
from $r_{o}/2M=5$.  The superposition can be seen in the interval $5<u/2M<20$;
it is small for $\alpha=1$, but large for $\alpha\neq 1$.
The third stage of the radiative process is the quasi-normal ringing of the black hole,
excited by the particle as it reaches the event horizon.
This phase of quasi-normal ringing is to be distinguished from the earlier
phase associated with the transmitted pulse; this new phase is insensitive to the choice
of initial data.

\begin{figure}[!t]
\vspace*{2.5in}
\special{hscale=31 vscale=31 hoffset=-13.0 voffset=190.0
         angle=-90.0 psfile=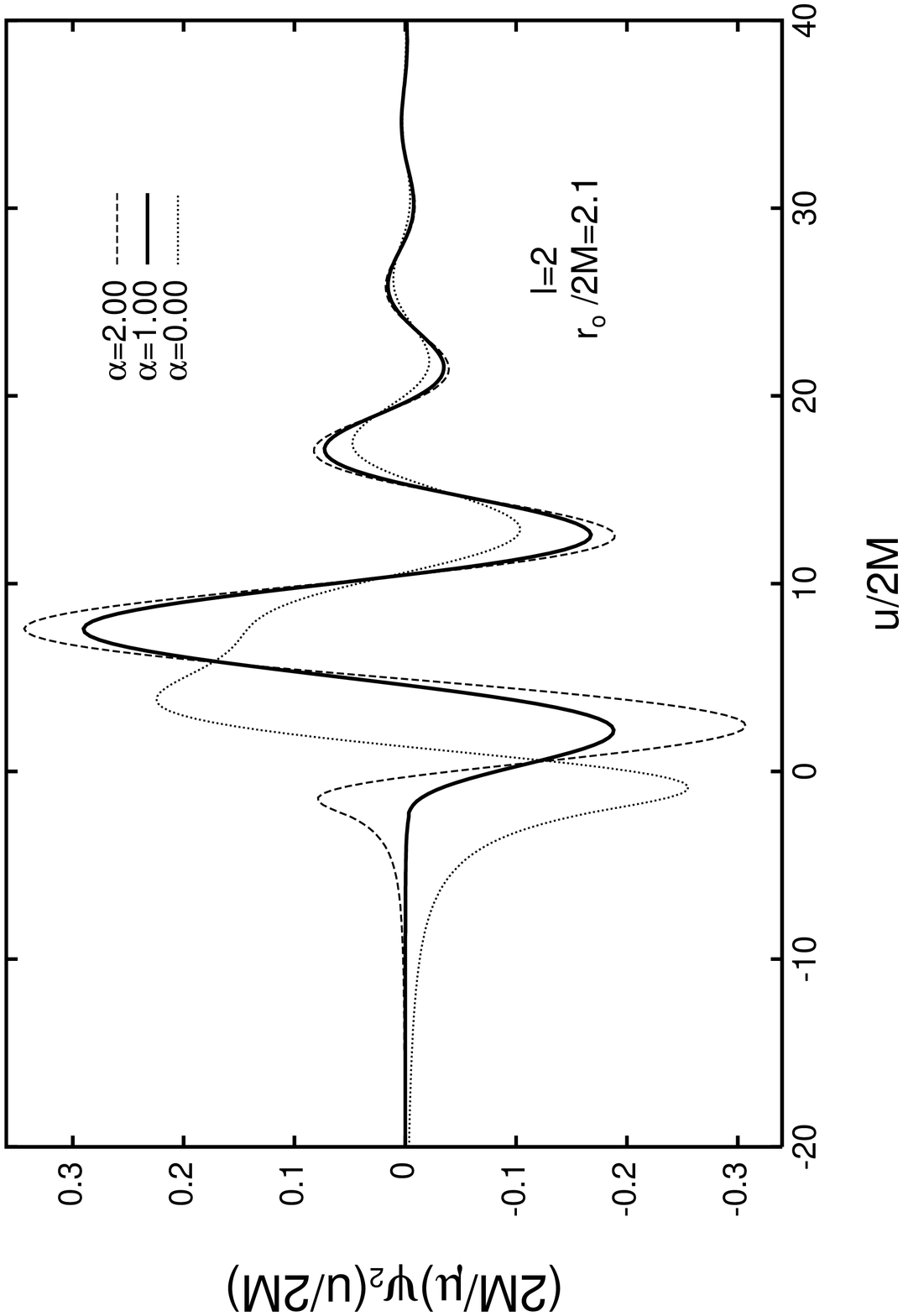}
\special{hscale=31 vscale=31 hoffset=225.0 voffset=190.0
         angle=-90.0 psfile=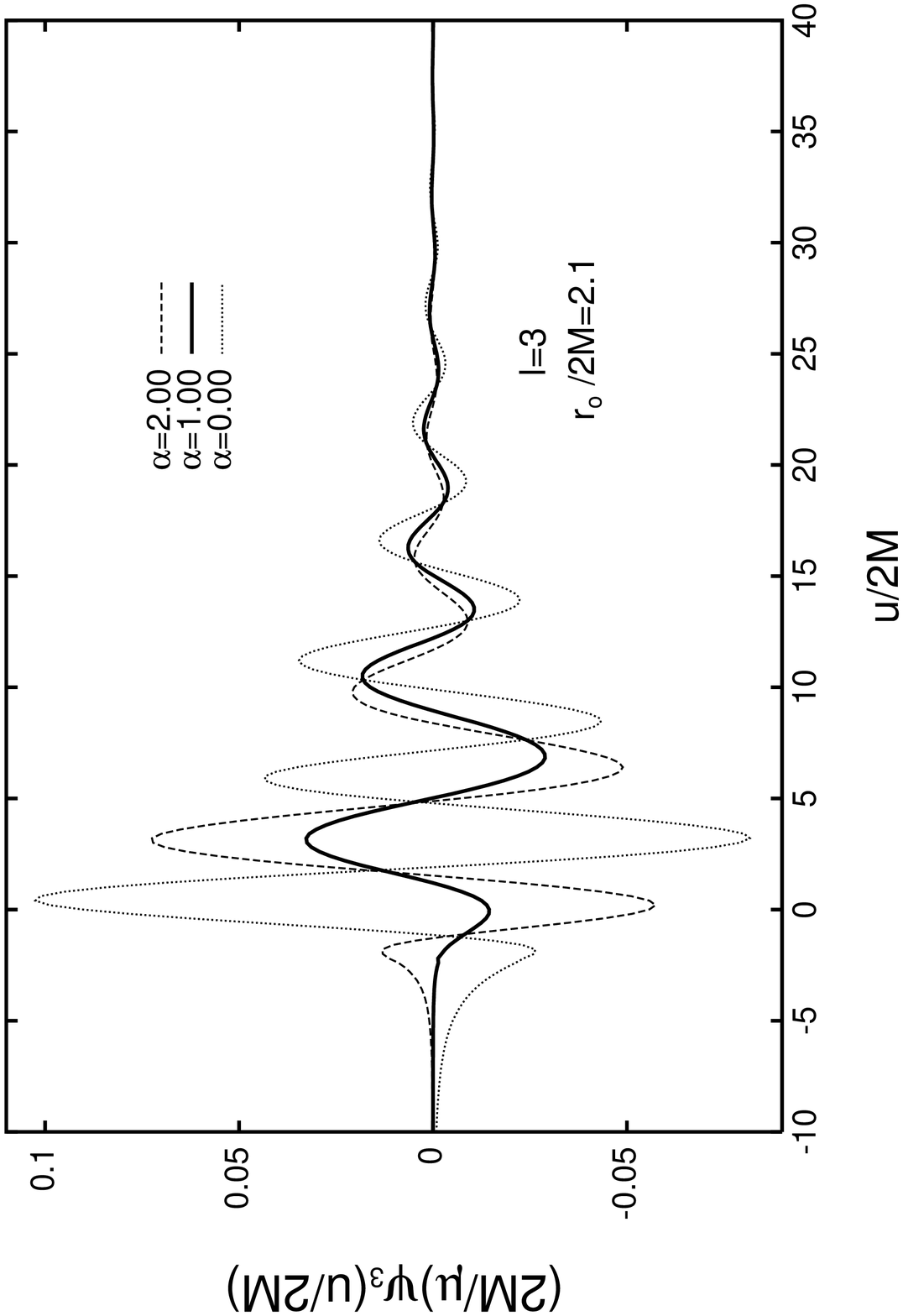}
\caption{The $l=2$ and 3 modes of the Zerilli-Moncrief function for a particle
falling from $r_{o}/2M=2.1$, for $\alpha=0$, 1,
and 2.  At intermediate times, the waveform is a superposition of the
backscattered ingoing pulse from the initial data, and acceleration
radiation from the particle.  The choice of initial data has an influence at
late times; it affects the amplitude and the phase
of the quasi-normal ringing.} \label{fig:Pr2p1}
\end{figure}
As $r_{o}$ is decreased further ($1.3\lesssim r_{o}/2M\lesssim 4$), the
waveforms become increasingly confused.
In this range of initial separations, the particle radiates strongly immediately,
and the reflected pulse of ingoing radiation does not reach infinity before
the radiation from the particle becomes significant.  The interference between
these two contributions to the waveform is fairly small,
because the particle radiates for a short time before passing through
the event horizon.  Figure~\ref{fig:Pr2p1} displays the $l=2$ and 3 modes of the
Zerilli-Moncrief function for infall from $r_{o}/2M=2.1$ for
$\alpha=0$, 1, and 2.  The interference between the backscattered ingoing
pulse and the radiation emitted by the particle is apparent, especially for
$l=2$ and $\alpha=0$.  A new phenomenon is observed for these values of
$r_{o}$: the stage of pure quasi-normal
ringing is now affected by the choice of initial data. The event horizon is
distorted both by the transmitted ingoing pulse and the particle, and these
factors act at roughly the same time.  Different choices of initial data will
therefore affect differently the amplitude and the phase of the quasi-normal
ringing.  This is displayed in Fig.~\ref{fig:Pr2p1}.

\begin{figure}[!t]
\vspace*{2.5in}
\special{hscale=31 vscale=31 hoffset=-13.0 voffset=190.0
         angle=-90.0 psfile=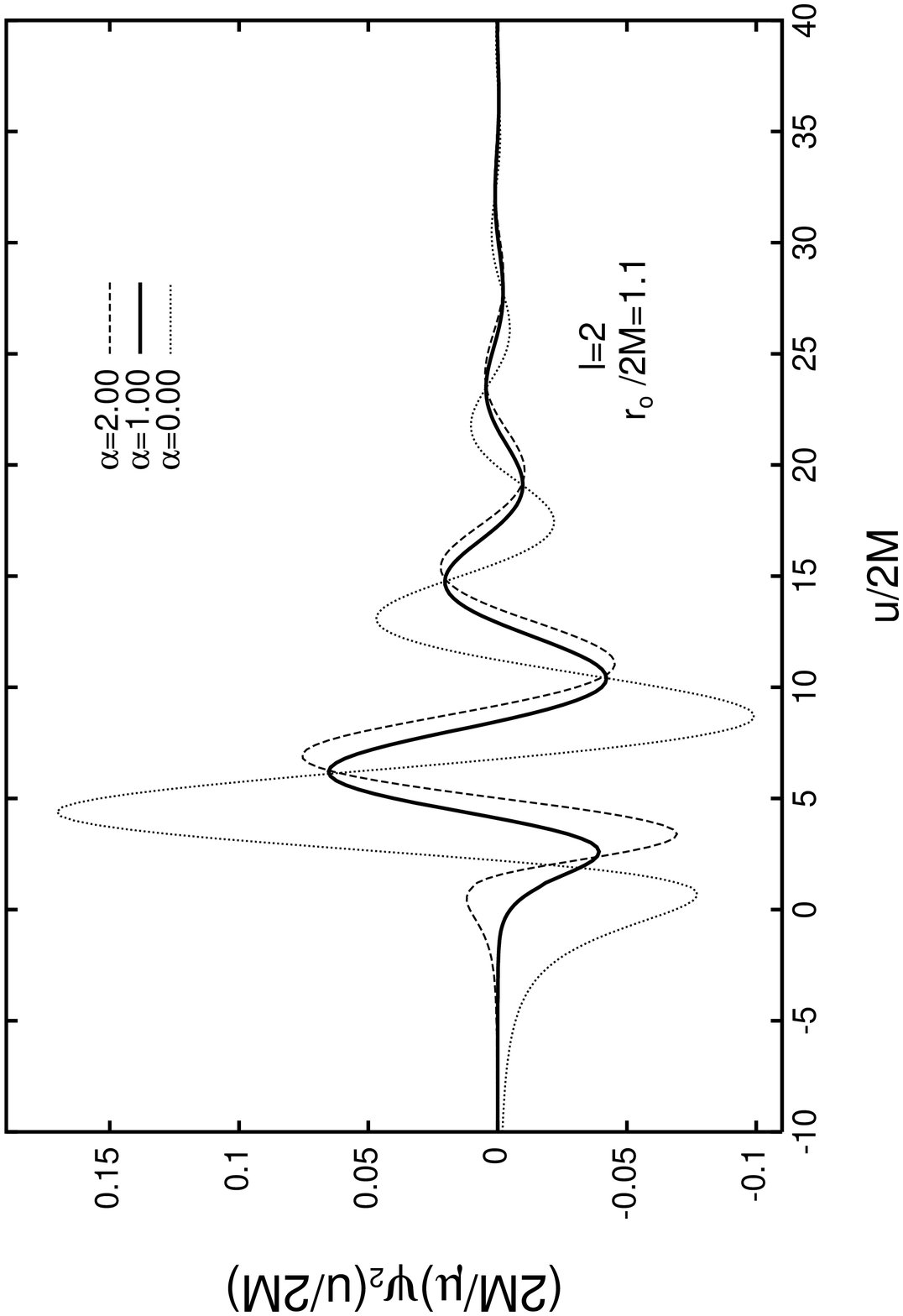}
\special{hscale=31 vscale=31 hoffset=225.0 voffset=190.0
         angle=-90.0 psfile=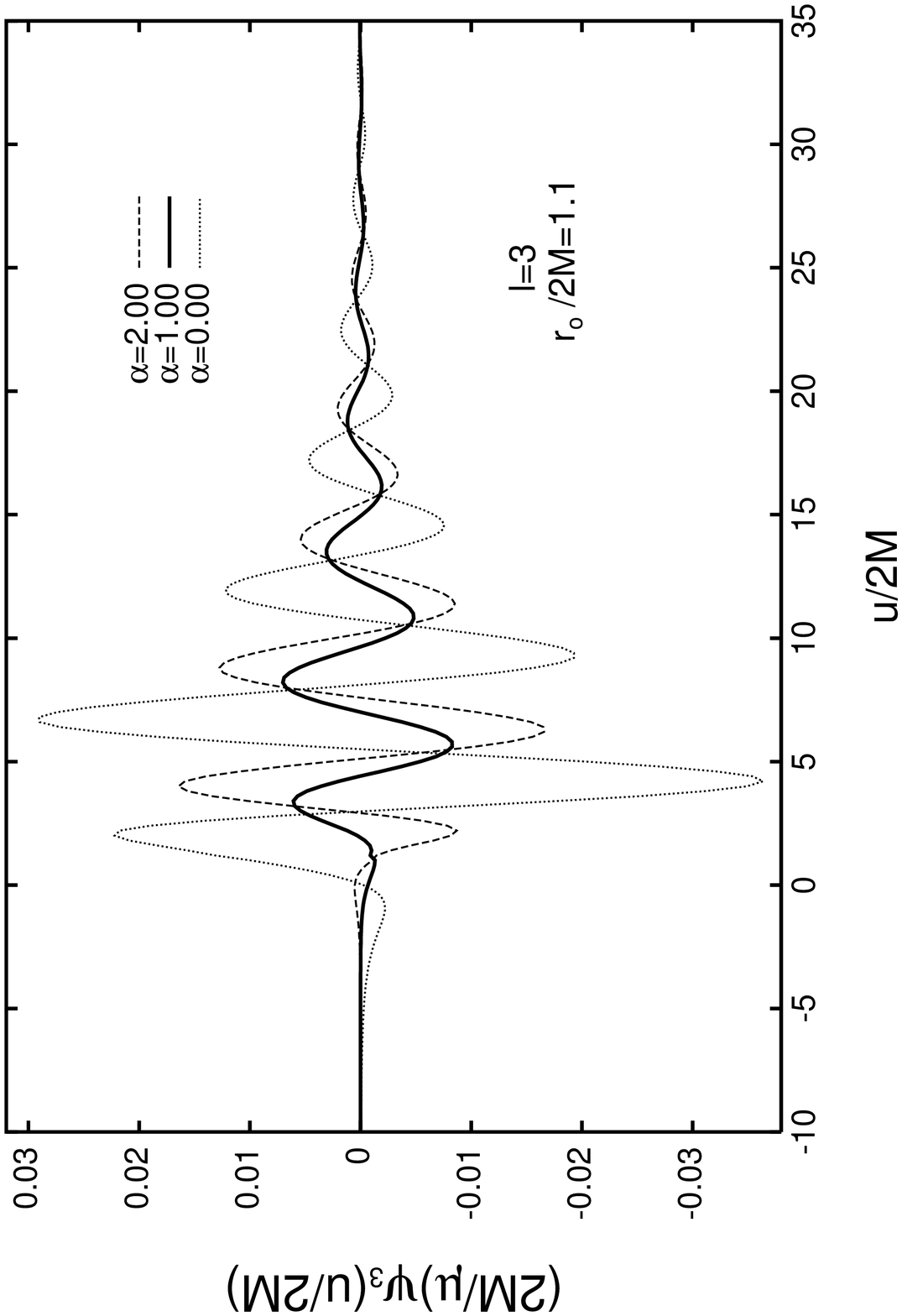}
\caption{The $l=2$ and 3 modes of the Zerilli-Moncrief function for a particle
falling from $r_{o}/2M=1.1$, for $\alpha=0$, 1, and 2.  For this very
small initial separation, the particle is quickly absorbed by the black hole
and cannot radiate much.  The initial value of $\psi(r,t)$ determines how strongly the
event horizon is distorted, and it therefore has a major impact on the quasi-normal
ringing phase of the radiation.} \label{fig:Pr1p1}
\end{figure}
These effects disappear when $r_{o}$ is moved past the potential
barrier ($1 < r_{o}/2M \lesssim 1.3$).  In such cases, most of the
outgoing pulse in the initial data is reflected by the potential barrier,
and does not register at infinity.
Instead, the pulses in the initial data and the radiation from the
particle work together to distort the event horizon.  For these very small
$r_{o}$, the quasi-normal ringing is the only feature that remains
in the waveforms.  The amplitude and the phase of the quasi-normal
ringing are very sensitive to the choice of initial data, because the information
about the initial distortion of the event horizon is entirely encoded in the initial data
(see Fig.~\ref{fig:Pr1p1} for the case $r_{o}/2M=1.1$).  The contribution from
the particle is minimal because it is almost immediately absorbed by the black hole.

\subsection{Power Spectra} \label{sec:spectrum}
\begin{figure}[!t]
\vspace*{4.8in}
\special{hscale=31 vscale=31 hoffset=-13.0 voffset=358.0
         angle=-90.0 psfile=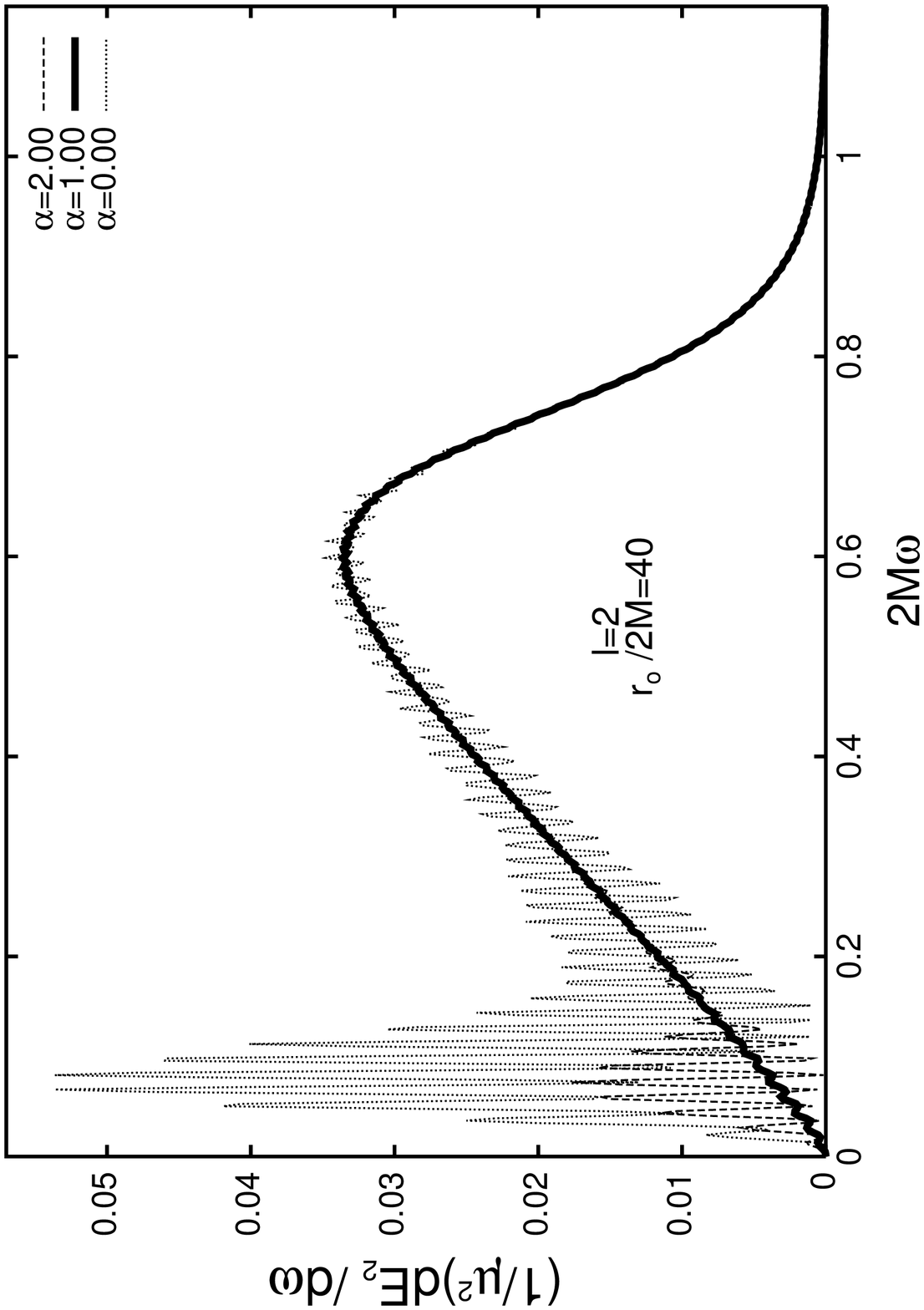}
\special{hscale=31 vscale=31 hoffset=225.0 voffset=358.0
         angle=-90.0 psfile=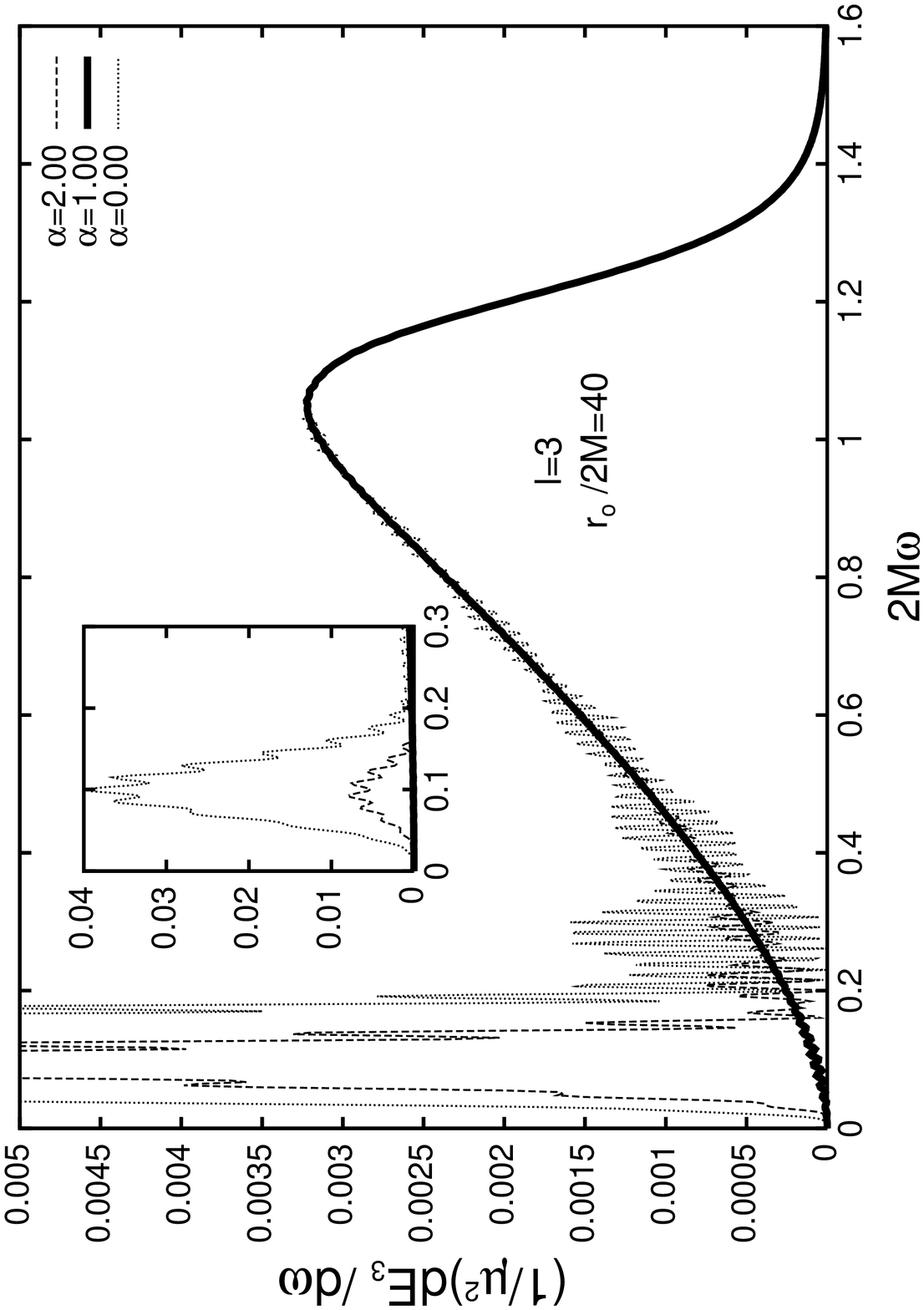}
\special{hscale=31 vscale=31 hoffset=-13.0 voffset=190.0
         angle=-90.0 psfile=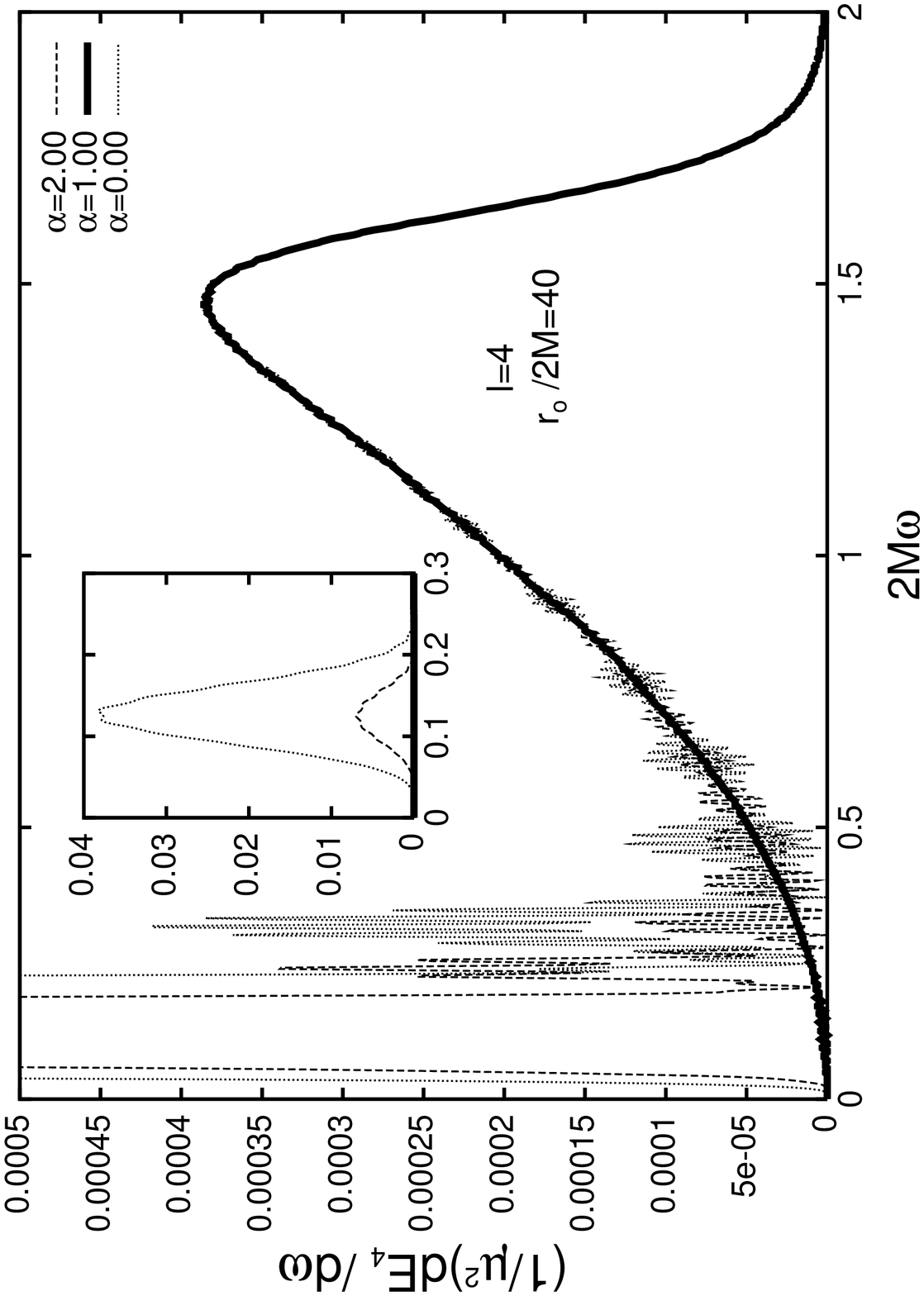}
\caption{Power spectra for the $l=2$, 3, and 4 modes of the Zerilli-Moncrief
function for $r_{o}/2M=40$, and $\alpha=0$, 1, and 2.  Varying $\alpha$ changes
the shape of the spectra at low frequencies, but the effect disappears at high
frequencies.  The low-frequency oscillations are more important for $\alpha=0$,
compared with $\alpha=1$ and 2, and $\alpha=2$ produces more oscillations
than $\alpha=1$.  The low-frequency part of
the spectrum is due mostly to the low-frequency gravitational waves contained in the
initial data.} \label{fig:Sr40}
\end{figure}
The power spectra, as calculated from Eq.~(\ref{eqn:spc}) with a fast Fourier
transform algorithm, tell a similar story.  For a given choice of $l$, $\alpha$, and $r_{o}$, the
total power spectrum is the sum of initial-data, particle, and
quasi-normal ringing contributions, but it includes interference between these contributions.

The interference is important when the pulses associated with the initial data, the acceleration
radiation, and the quasi-normal ringing contain overlapping frequencies.
Typical frequencies are $\omega_{o} \sim r_{o}^{-1}$ for the initial-data radiation,
$\omega_{p} \sim r^{-1}_{p}$ for the acceleration radiation, and
$\omega_{QN}\sim M^{-1}$ for the quasi-normal ringing.
As was mentioned in Sec.~\ref{sec:waves}, the particle emits mostly in the
strong-field region of the spacetime ($\omega_{p}\sim M^{-1}$) and
consequently, the interference between acceleration radiation and quasi-normal
ringing is always present, while interference with
the initial-data contribution is important only when $r_{o}$ is not much larger than $2M$.
In other words, interference effects involving the initial-data pulses are important
when the particle is released in the strong-field region.  This picture
is somewhat simplistic, but it serves as a useful guide to
determine when interference effects become important.  Our numerical results are
consistent with this picture.

For large initial separations ($r_{o}/2M \gg 1$), we expect the energy spectrum to be the direct
sum of the powers in the initial-data and particle contributions, and in the quasi-normal ringing,
without much interference.  This is confirmed in
Fig.~\ref{fig:Sr40}, where the spectra for the $l=2$, 3, and 4 modes of the
Zerilli-Moncrief function for $r_{o}/2M=40$, and $\alpha=0$, 1, 2, are
presented.  At low frequencies, the initial data manifests itself as a strong
pulse which dominates the spectrum.  At higher frequencies, acceleration radiation
and quasi-normal ringing dominate the spectrum; this part of the spectrum is easy
to recognize, as it does not change when $\alpha$ is varied.
As we move away from conformal flatness ($\alpha=1$), the influence of the initial data
spreads into higher frequencies.
This can be seen as oscillations in the spectrum (cf. the cases $\alpha=0$ and 2 in Fig.~\ref{fig:Sr40}).
These oscillations are also present for $\alpha=1$, but
increasing or decreasing the value of $\alpha$ increases their
amplitude and the extent by which they spill into higher
frequencies.  In general, a choice of initial data with $\alpha <1$, instead
of $\alpha\geq 1$, produces larger oscillations, and the effect extends
to higher frequencies. (The oscillations are the smallest when $\alpha=1$.)

\begin{figure}[!t]
\vspace*{4.8in}
\special{hscale=31 vscale=31 hoffset=-13.0 voffset=358.0
         angle=-90.0 psfile=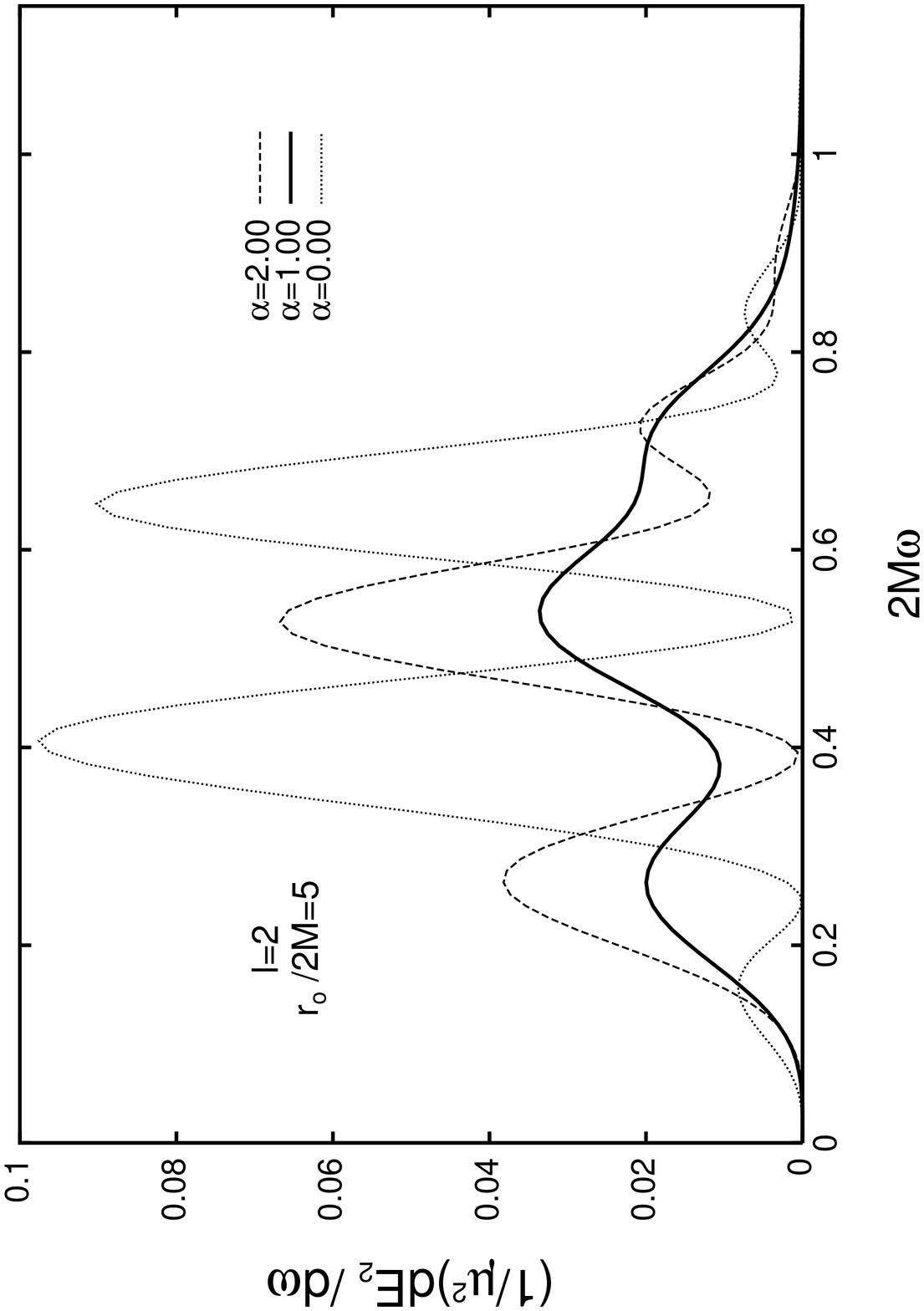}
\special{hscale=31 vscale=31 hoffset=225.0 voffset=358.0
         angle=-90.0 psfile=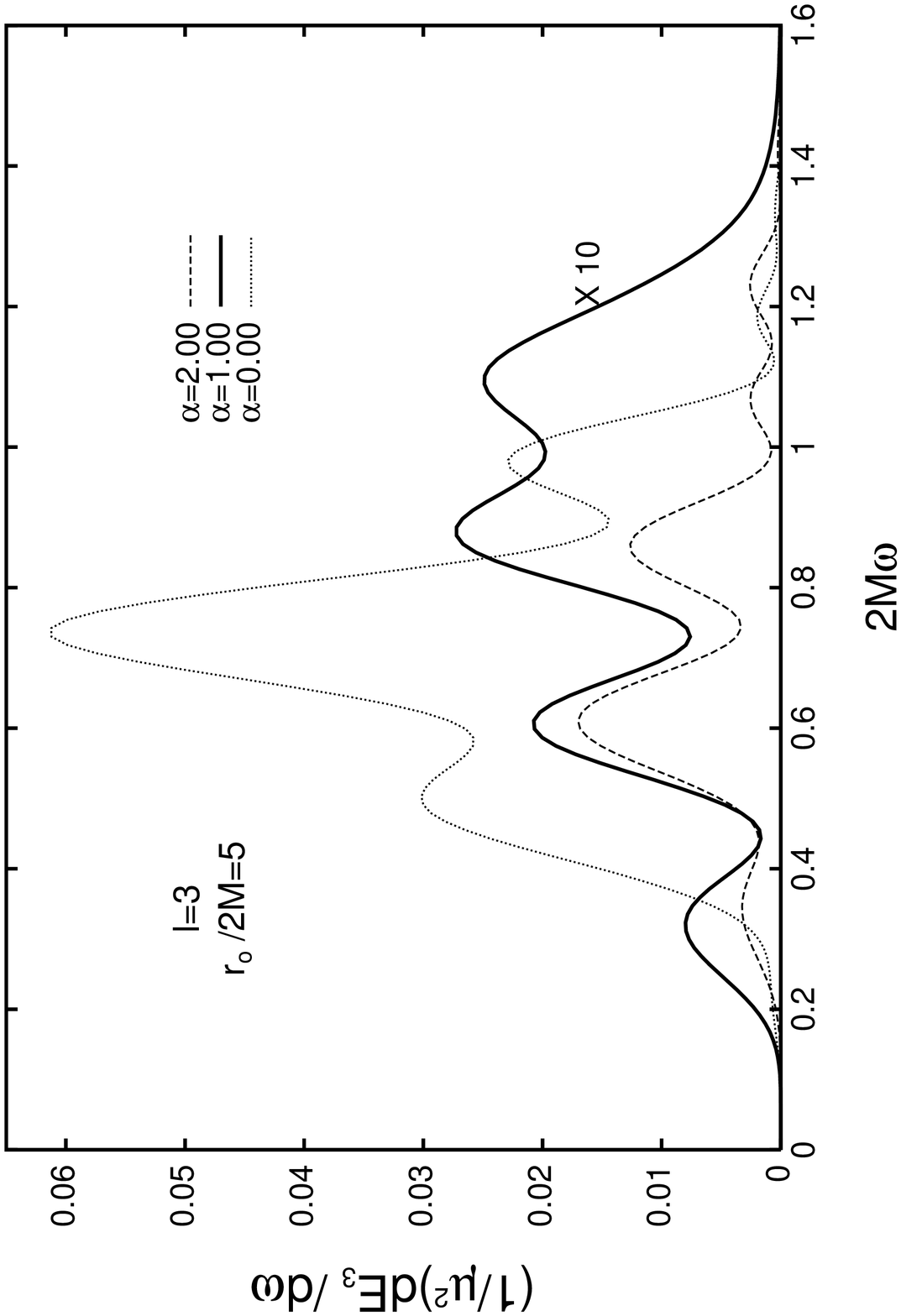}
\special{hscale=31 vscale=31 hoffset=-13.0 voffset=190.0
         angle=-90.0 psfile=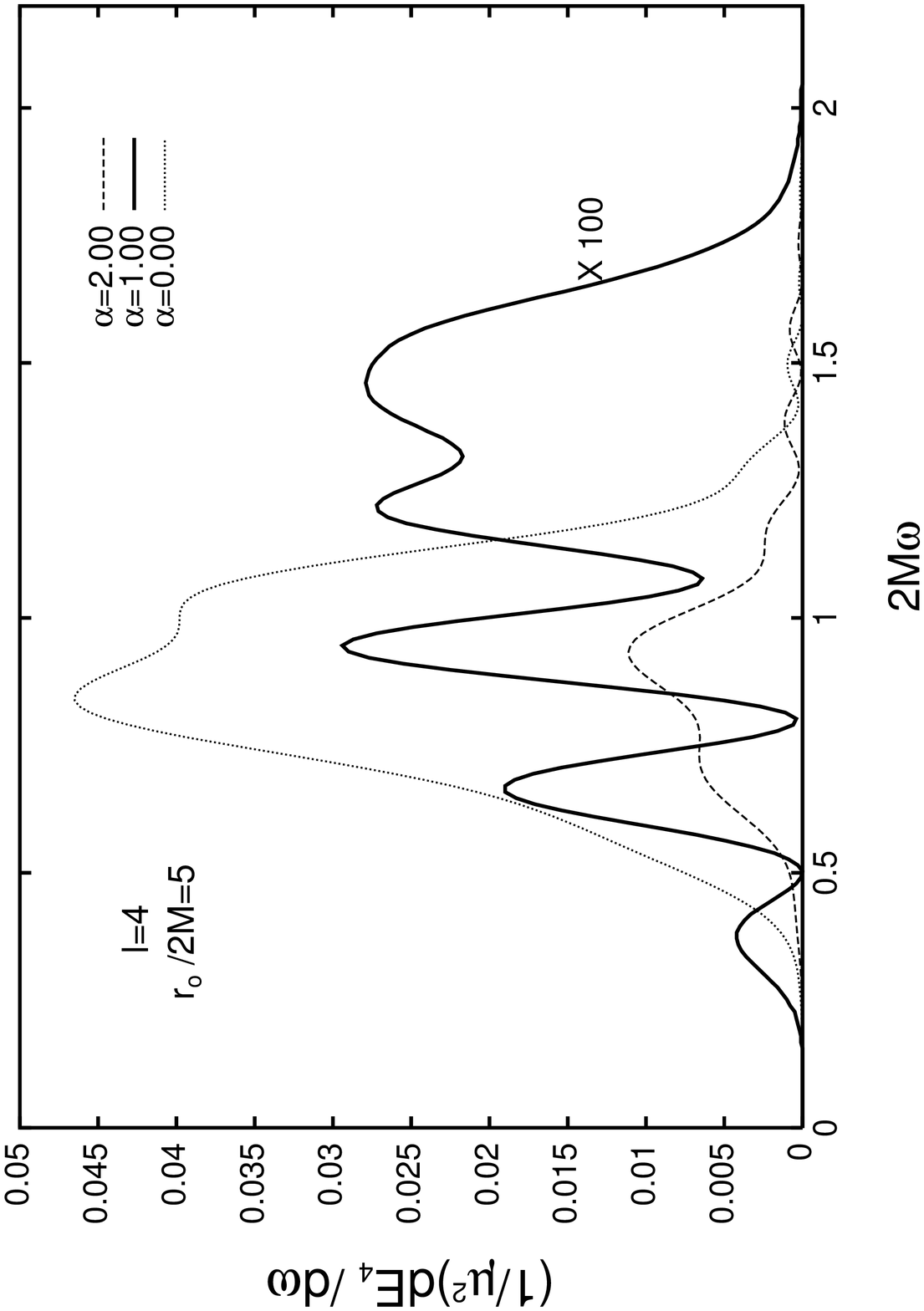}
\caption{Power spectra for the $l=2$, 3, and 4 modes of the Zerilli-Moncrief function for
infall from $r_{o}/2M=5$, and $\alpha=0$, 1, and 2.  Interference between the initial-data
pulses, the acceleration radiation, and the quasi-normal ringing plays a
crucial role in determining the shape of the
spectra.  For $\alpha=0$ and $l=2$, the spectrum has two maxima at
$2M\omega=0.4065$ and $2M\omega=0.6455$, frequencies at which the spectra for
$\alpha=1$ and 2 have a local minimum.  Similarly, the maximum at $2M\omega=0.5364$
for $\alpha=1$ and 2 is replaced by a minimum when $\alpha=0$.  This is indicative
of strong interference effects.  The labels ``$\times 10$'' and ``$\times 100$''
indicate the amount by which the amplitude of these two spectra were multiplied to be
presented in the same figure.} \label{fig:Sr5}
\end{figure}
We have seen in the previous section that for intermediate values of $r_{o}$
($4\lesssim r_{o}/2M <10$),
the choice of initial data affects the shape of the waveforms up to times
where acceleration radiation starts to dominate the radiative process.
For these initial separations, interference effects are important.
This is confirmed by our numerical simulations of infall from $r_{o}/2M=5$.
The spectra, for this value of $r_{o}$, are displayed in Fig.~\ref{fig:Sr5}
for $l=2$, 3, and 4, and $\alpha=0$, 1, and 2.  For $l=2$, initial data with
$\alpha>1$ tend to amplify the features apparent for $\alpha=1$
(see the case $\alpha=2$ in Fig.~\ref{fig:Sr5}), while decreasing the value
of $\alpha$ changes the location of the maxima; local minima
appear in the spectrum where maxima were seen for $\alpha=1$ (see $\alpha=0$
in Fig.~\ref{fig:Sr5}).  For $l=3$ and $4$ and $\alpha>1$ ($\alpha<1$), we observe a similar amplification (attenuation)
of the features present for $\alpha=1$, but the effect is much weaker than for $l=2$.

\begin{figure}[!t]
\vspace*{2.5in}
\special{hscale=31 vscale=31 hoffset=-13.0 voffset=190.0
         angle=-90.0 psfile=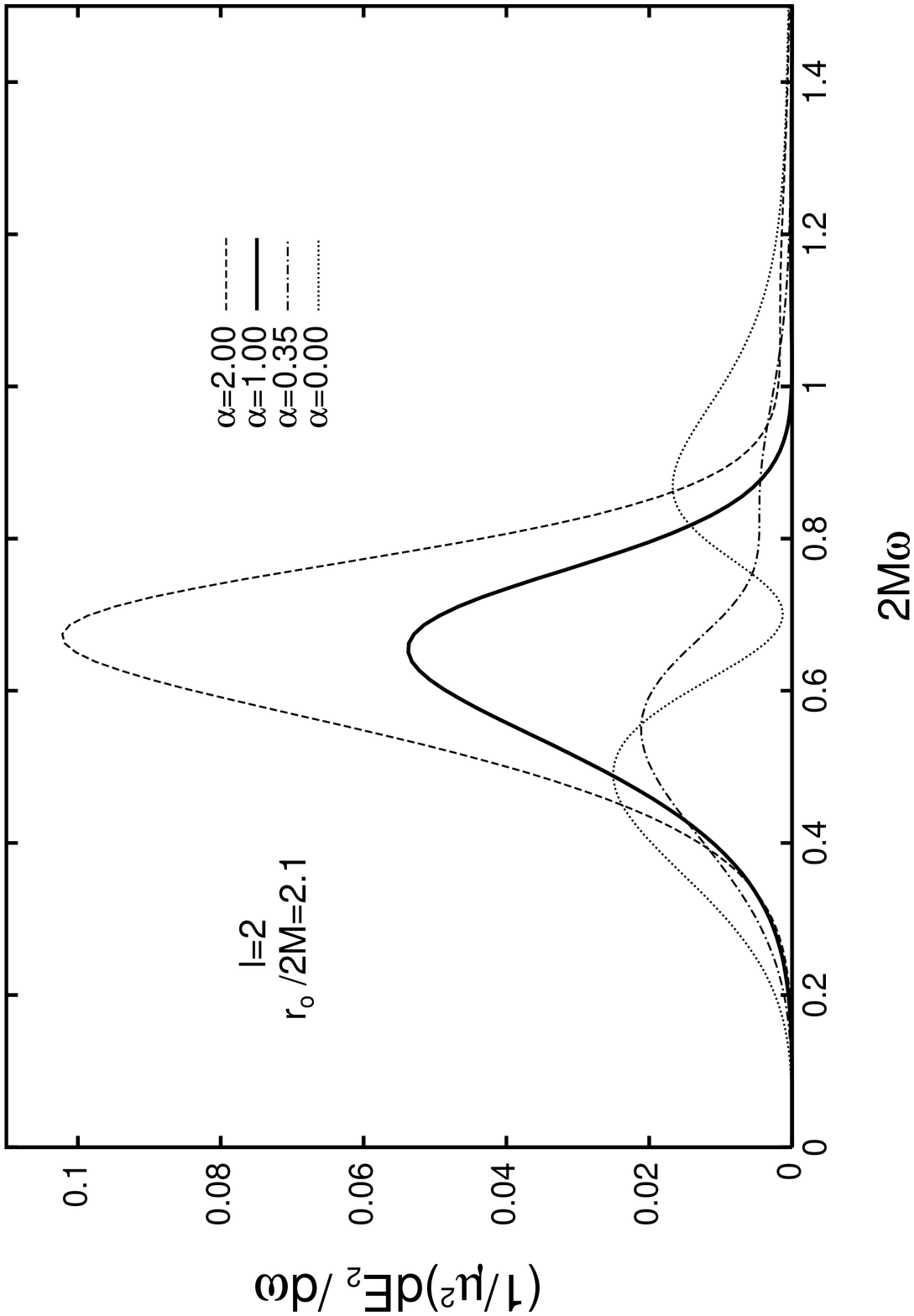}
\special{hscale=31 vscale=31 hoffset=225.0 voffset=190.0
         angle=-90.0 psfile=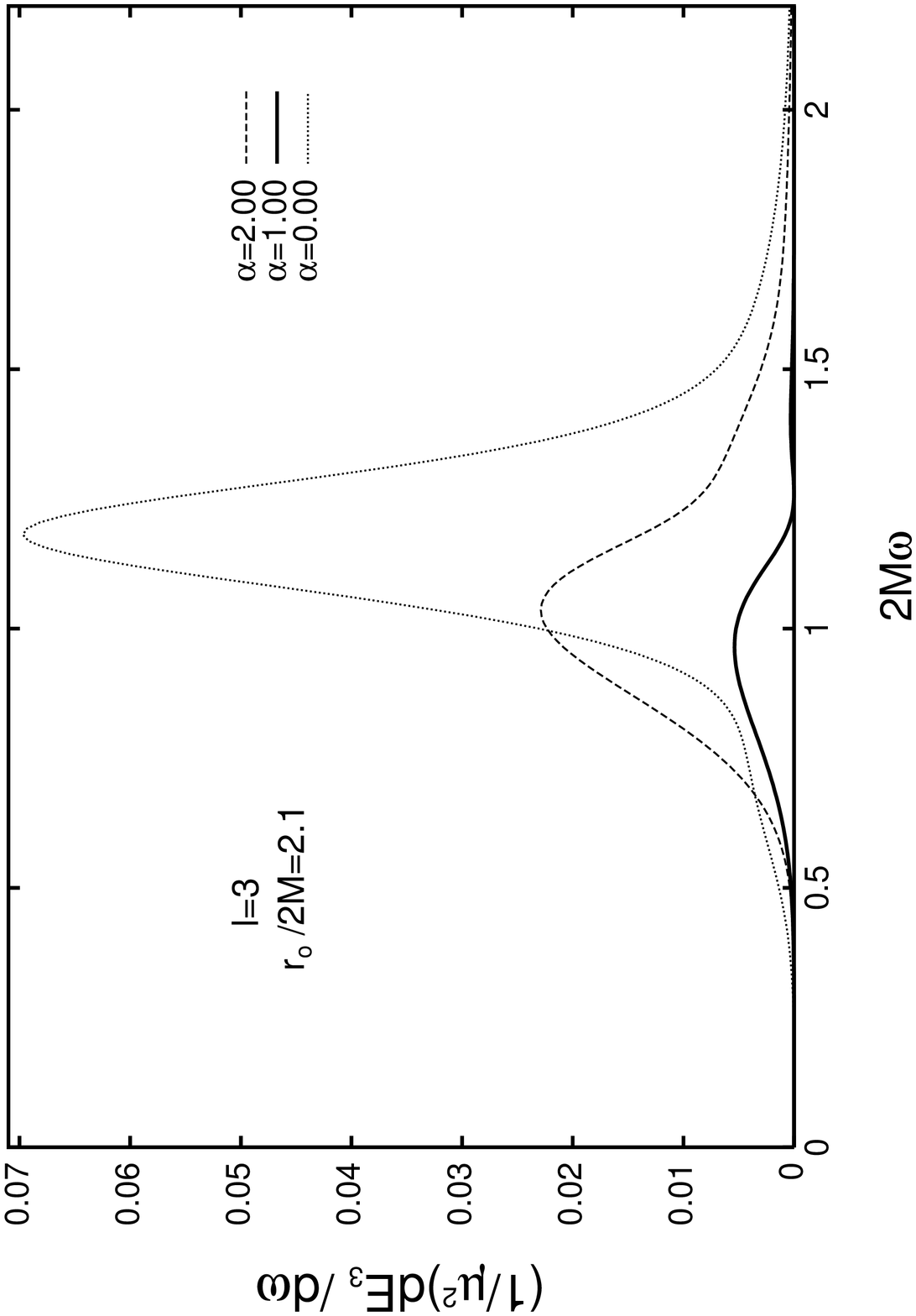}
\caption{Power spectra for the $l=2$ and 3 modes of the Zerilli-Moncrief function
for infall from $r_{o}/2M=2.1$, and $\alpha=0$, 1, and 2.  For $l=2$, we also
display $\alpha=0.35$, to show the smooth transition between
the single-maximum and two-maxima regimes described in the text.  For such a small
separation, acceleration radiation is small, and the interference
effects are not very pronounced.  They still, however, play
an important role in determining the shape of the spectra.  The spectrum for $l=2$
and $\alpha=0$ has a shape that indicates strong interference effects between
initial-data, particle, and quasi-normal ringing contributions: the
single maximum at $2M\omega=0.6553$, seen for $\alpha=1$ and 2, is replaced by
two maxima at $2M\omega=0.4915$ and $2M\omega=0.8697$.} \label{fig:Sr2p1}
\end{figure}
As $r_{o}$ is taken closer to the potential barrier ($1.3 \lesssim r_{o}/2M \lesssim 4$),
the interference
becomes less important.  In Sec.~\ref{sec:waves}, we showed that for $r_{o}$ close
to the potential barrier, the particle does not radiate strongly before passing
through the event horizon, and the interference is small because of
the small amount of
acceleration radiation.  In Fig.~\ref{fig:Sr2p1} we present the spectra for
infall from $r_{o}/2M=2.1$, $\alpha=0$, 1, and 2, and $l=2$, and 3.
Although interference effects are not as important as for infall from $r_{o}/2M=5$,
they still play a role in determining the shape of the spectrum, as
can be seen for the case $\alpha=0$.  In this case, the spectrum has two maxima
and a single minimum; this minimum occurs close to the fundamental quasi-normal frequency,
which suggests a strong destructive interference between initial-data-excited and
particle-excited quasi-normal ringing.  This is different from the cases
$\alpha=1$ and 2, for which the spectra contain a single peak, indicating that most
of the energy is radiated into quasi-normal modes.

Interference effects, such as the ones shown in Fig.~\ref{fig:Sr2p1}, have been
observed previously by Lousto\cite{Lousto}, but in a different context.
Instead of evolving the perturbed Misner solution, appropriate
for a head-on collision in perturbation theory, Lousto chose to use the full Misner
solution\cite{Misner} as initial data for the Zerilli-Moncrief function.  He then
evolved these initial data using Eq.~(\ref{eqn:wave}).  In this case,
the interference is due to the non-linear nature of the initial data he evolved.

\begin{figure}[!t]
\vspace*{2.5in}
\special{hscale=31 vscale=31 hoffset=-13.0 voffset=190.0
         angle=-90.0 psfile=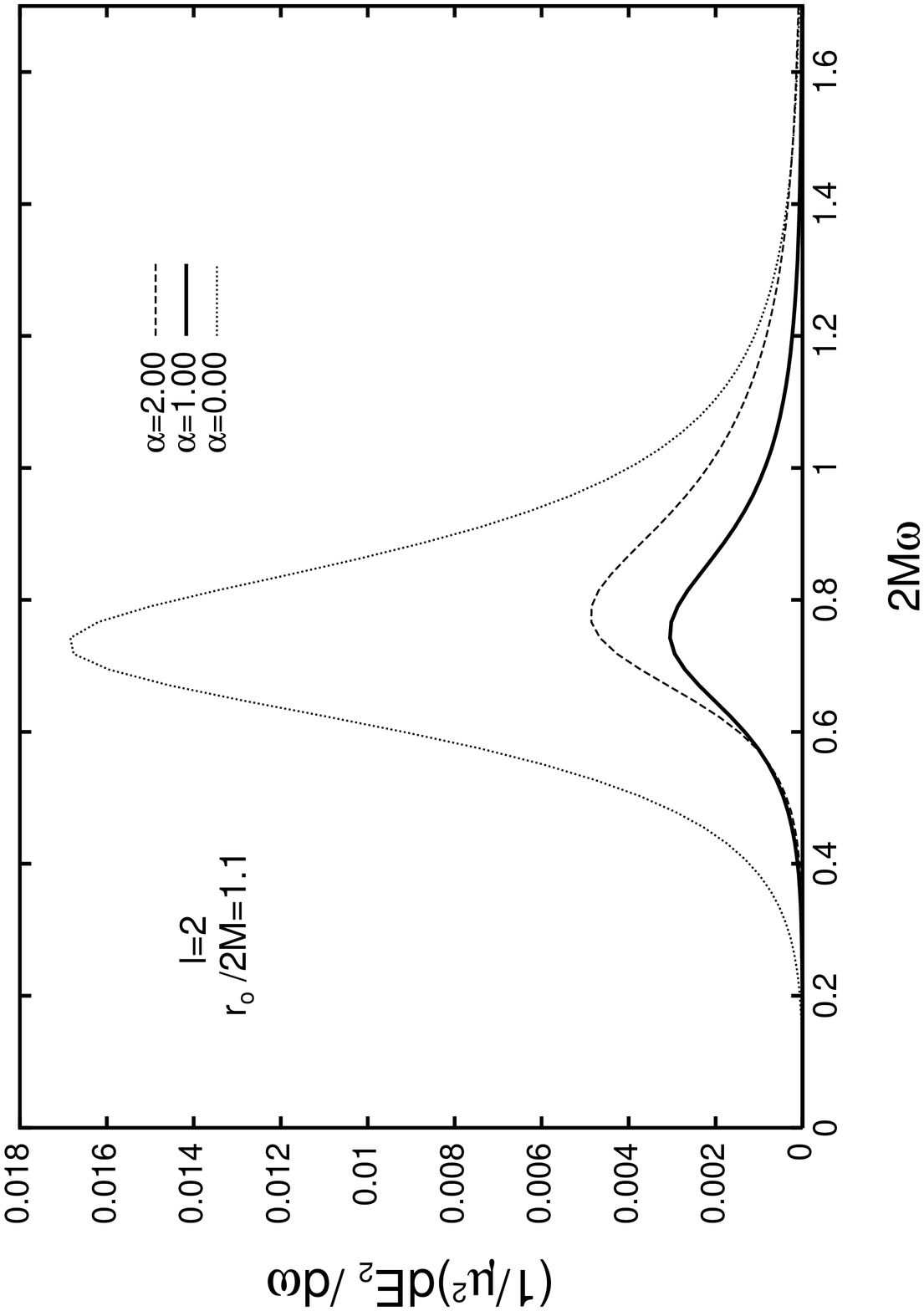}
\special{hscale=31 vscale=31 hoffset=225.0 voffset=190.0
         angle=-90.0 psfile=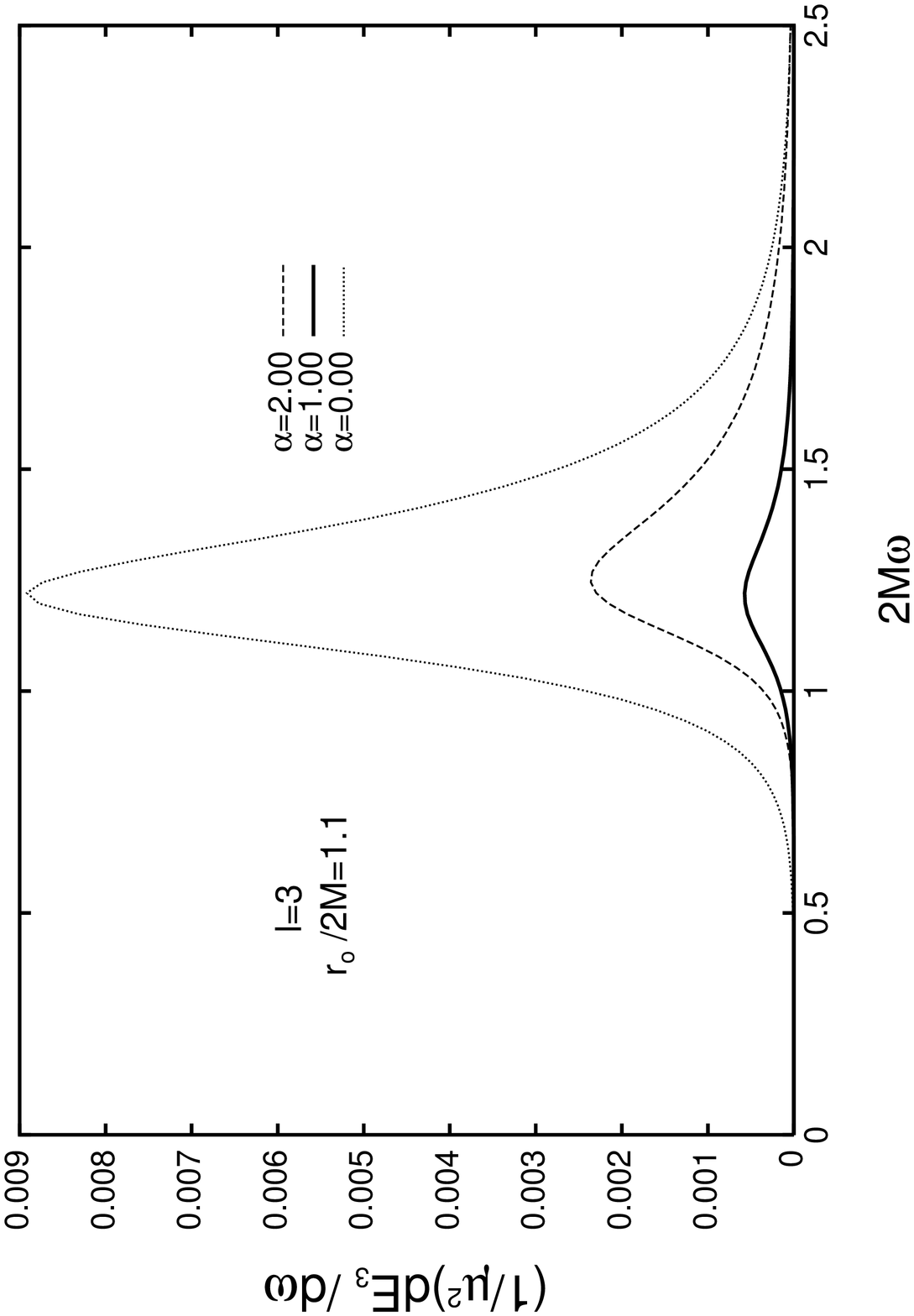}
\caption{Power spectra for the $l=2$ and 3 modes of the Zerilli-Moncrief function
for infall from $r_{o}/2M=1.1$, and $\alpha=0$, 1, and 2.  For
$r_{o}$ well inside the potential barrier, acceleration radiation is negligible
since the particle is absorbed by the black hole immediately.  The initial
data distorts the event horizon, which becomes dynamical and starts
radiating at its quasi-normal frequencies.
The amplitude of the quasi-normal ringing is determined by the strength of the
tidal distortion exerted on the event horizon by the initial data.  For the cases
displayed, the lowest amplitude is obtained for $\alpha=1$, while the highest is obtained
for $\alpha=0$.} \label{fig:Sr1p1}
\end{figure}
Finally, when $r_{o}$ is chosen well inside the potential
barrier ($1 < r_{o}/2M \lesssim 1.3$), the interference effects mentioned previously disappear.
For small $r_{o}$, the magnitude of the tidal distortion applied to the event
horizon is affected only by the choice of initial data.  Since the radiation
is then dominated by quasi-normal ringing created by the initial tidal distortion of the black hole,
the spectra contain a single peak, and its position is independent of $\alpha$.  The choice of
initial data affects only the amplitude of the quasi-normal ringing, and this is
reflected in the amplitude of the spectra at
the fundamental quasi-normal frequency.  For infall from $r_{o}/2M=1.1$, displayed in
Fig.~\ref{fig:Sr1p1} for $l=2$ and 3, and $\alpha=0$, 1, and 2, the lowest
amplitude occurs for $\alpha=1$, while the highest occurs for $\alpha=0$.

\subsection{Total Energy Radiated} \label{sec:energy}
In this subsection we calculate the total energy radiated as a function of $\alpha$
for infall from $r_{o}/2M=40$, and as a
function of $r_{o}$ for six selected values of $\alpha$.  We also tabulate the
total energy radiated (the sum of the $l=2$, 3 and 4 modes) for selected values
of $r_{o}$ and $\alpha$.  The radiated energy was calculated using two different
methods: direct numerical integration of Eq.~(\ref{eqn:power}), and numerical
integration of Eq.~(\ref{eqn:spc}).  In this way, the accuracy of the
algorithms used could be tested.  The results were found to agree to better than
$1\%$.  For comparison, we include in our figures the energy radiated by a particle
falling from infinity, as calculated by Davis, Ruffini, Press, and
Price (DRPP)\cite{DRPP}.  The
DRPP result for the energy radiated in the $l=2$ mode is $(2M/\mu^2) E_{2}=1.84\times 10^{-2}$.

\begin{figure}[!ht]
\vspace*{2.5in}
\special{hscale=31 vscale=31 hoffset=-13.0 voffset=190.0
         angle=-90.0 psfile=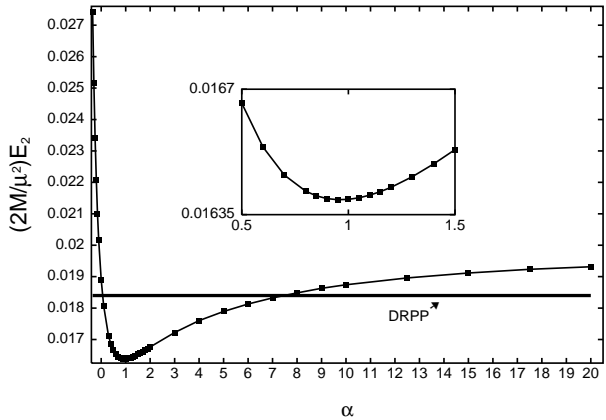}
\caption{The total energy radiated in the $l=2$ mode as a function of $\alpha$ for a
particle falling in from $r_{o}/2M=40$.  Outside of the range $0 < \alpha < 7.5$, the
total energy radiated exceeds the DRPP result.  The energy is minimized for $\alpha\approx 1$.
}\label{fig:enrg40a}
\end{figure}

\begin{figure}[!ht]
\vspace*{2.7in}
\special{hscale=31 vscale=31 hoffset=-13.0 voffset=190.0
         angle=-90.0 psfile=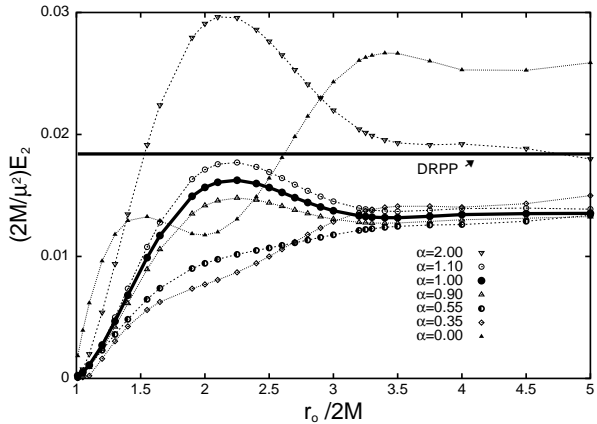}
\caption{The total energy radiated by the $l=2$ mode as a function of $r_{o}$.
Displayed are six curves corresponding to the values $\alpha=0$, 0.55, 0.9, 1, 1.1, and 2.
For these small values of $r_{o}$, $\alpha=1$ no longer minimizes the energy radiated.
Instead, the value of $\alpha$ that minimizes the energy changes
with $r_{o}$.  This reflects the interference between the initial data, the particle,
and the quasi-normal ringing contributions to the radiation.  We were unable to find a
single value of $\alpha$ that minimizes the energy in the interval
$1<r_{o}/2M \leq 5$.}\label{fig:enrg2r}
\end{figure}
We have seen that for large initial separations, the different contributions to
the gravitational radiation do not interfere.  The energy emitted in each
multipole is then the direct sum of the energy
contained in the initial-data pulses, the energy emitted in the
acceleration radiation, and the energy radiated in the quasi-normal modes.
In Sec.~\ref{sec:waves} we showed that all choices of initial data result in
two pulses at early times: the outgoing pulse and the backscattered ingoing
pulse. These pulses have the smallest amplitude when $\alpha=1$.  Hence, for
large initial separations, the waveforms obtained with $\alpha=1$ are
the ones that carry the least amount of energy to infinity.

The total energy radiated in the $l=2$ mode is shown in Fig.~\ref{fig:enrg40a} 
as a function of $\alpha$, for $r_{o}/2M=40$.  
We see that the energy is minimized for $\alpha \approx 1$, and that it exceeds the 
DRPP result if $\alpha$ is outside the interval 
$0\lesssim \alpha \lesssim 7.5$.  For such a large value of the initial radius, 
we confirm the general belief that the radiated energy is minimized if the 
initial hypersurface is conformally flat\cite{York1,Baumgarte}. 

\begin{table}
\caption{\label{tab:enrg}Total energy radiated (in units of $2M/\mu^2$) for the $l=2$, 3, and 4 modes, 
and $\alpha=0.75$, 0.8, 0.85, 0.9, and 1.0.  The sum of the first three 
multipole moments is denoted by $\bf{E}$. For infall from short and 
intermediate distances, the minimum in $\bf{E}$ is achieved for $\alpha<1$.  }
\begin{ruledtabular}
\begin{tabular}{ccccccc} \\ \hline
$r_{o}/2M$ & $l$ & $\alpha$=0.75 & $\alpha=$0.8 & $\alpha$=0.85 & $\alpha$=0.9 & $\alpha$=1.0  \\
\hline \hline
10.0 &2& 1.467e-2 & 1.464e-2 & 1.463e-2 & 1.464e-2 & 1.469e-2\\
     &3& 1.961e-3 & 1.876e-3 & 1.820e-3 & 1.789e-3 & 1.790e-3\\
     &4& 4.835e-4 & 3.778e-4 & 3.045e-4 & 2.595e-4 & 2.406e-4\\
     &$\bf{E}$& \bf{1.711e-2}& \bf{1.690e-2} & \bf{1.675e-2} & \bf{1.669e-2} & \bf{1.672e-2}\\
5.0  &2& 1.305e-2 & 1.309e-2 & 1.316e-2 & 1.326e-2 & 1.353e-2\\
     &3& 1.725e-3 & 1.610e-3 & 1.546e-3 & 1.529e-3 & 1.606e-3\\
     &4& 5.600e-4 & 3.883e-4 & 2.761e-4 & 2.156e-4 & 2.249e-4\\
     &$\bf{E}$& \bf{1.534e-2} & \bf{1.509e-2} & \bf{1.498e-2} & \bf{1.500e-2} & \bf{1.536e-2}\\
3.0  &2& 1.221e-2 & 1.246e-2 & 1.274e-2 & 1.305e-2 & 1.374e-2\\
     &3& 1.359e-3 & 1.307e-3 & 1.323e-3 & 1.398e-3 & 1.693e-3\\
     &4& 5.285e-4 & 3.298e-4 & 2.147e-4 & 1.721e-4 & 2.681e-4\\
     &$\bf{E}$& \bf{1.410e-2} & \bf{1.410e-2} & \bf{1.428e-2} & \bf{1.462e-2} & \bf{1.570e-2}\\
2.1  &2& 1.238e-2 & 1.310e-2 & 1.383e-2 & 1.457e-2 & 1.608e-2\\
     &3& 1.270e-3 & 1.325e-3 & 1.446e-3 & 1.625e-3 & 2.120e-3\\
     &4& 4.134e-4 & 2.515e-4 & 1.757e-4 & 1.748e-4 & 3.598e-4\\
     &$\bf{E}$& \bf{1.406e-2} & \bf{1.468e-2} & \bf{1.545e-2} & \bf{1.637e-2} & \bf{1.856e-2}\\
1.1  &2& 9.158e-4 & 9.068e-4 & 9.066e-4 & 9.144e-4 & 1.099e-3\\
     &3& 2.017e-4 & 1.892e-4 & 1.854e-4 & 1.894e-4 & 2.250e-4\\
     &4& 2.286e-4 & 5.415e-5 & 4.654e-5 & 4.529e-5 & 5.966e-5\\
     &$\bf{E}$& \bf{1.346e-3} & \bf{1.150e-3} & \bf{1.139e-3} & \bf{1.149e-3} & \bf{1.384e-3}\\
\end{tabular}
\end{ruledtabular}
\end{table}
In Fig.~\ref{fig:enrg2r} we plot the total energy radiated in the $l=2$ mode 
as a function of $r_{o}$ for selected values of $\alpha$.  In the range $1<r_{o}/2M\leq 5$, 
the energy is no longer minimized by the choice $\alpha=1$.  Instead, we find that 
in the interval $2.7 < r_{o}/2M \leq 5$, the energy is minimized when 
$\alpha=0.55$, while in the interval $1 \leq r_{o}/2M \leq 2.7$, the minimum is achieved 
when $\alpha=0.35$.  (These values are approximate, as we found it
difficult in practice to locate the true minimum in the energy for a given $r_{o}$ and
$l$.)  We also find that for a given $r_{o}$, no value of $\alpha$ minimizes the
energy radiated for all modes.  For example, for infall from $r_{o}/2M=5$ there is a minimum in
the energy for the $l=3$ and 4 modes when $\alpha=0.9$, but this value of $\alpha$ does not
minimize the energy in the $l=2$ mode (see Table~\ref{tab:enrg}).

Although we are unable to find a single value of $\alpha$ that minimizes
the energy radiated for all $l$ and over the whole interval $1<r_{o}/2M\leq 5$,
we see that in the cases considered, the total energy radiated is
never minimized by choosing initial data with $\alpha >1$.  Instead,
for infall from a small $r_{o}$, it is minimized by a choice of $\alpha<1$.   This
is due to strong interference effects for initial data with $\alpha<1$ when
$r_{o}$ in the strong-field region of the spacetime; this interference was discussed
in Secs.~\ref{sec:waves} and \ref{sec:spectrum}.

In Table~\ref{tab:enrg} we display the total energy radiated in the $l=2$, 3, and 4
modes of the  Zerilli-Moncrief function, for selected values of
$\alpha$ and $r_{o}$.  In general, the features presented previously for the case
$l=2$ are also present for higher multipole moments:  the total energy radiated in a given
multipole is not minimized by the conformally-flat choice of initial data
($\alpha=1$), but by some $\alpha<1$.
We find that in some cases, the interference effects become so important
that the energy radiated increases with the multipole order $l$, e.g.
$E_{2}<E_{3}<E_{4}$.  This increase in energy with increasing $l$ contradicts our
slow-motion
expectation, according to which the energy radiated should decrease with increasing
multipole order.  A typical example of this
phenomenon is displayed in Table~\ref{tab:enrg} for $r_{o}/2M=1.1$ and
$\alpha=0.75$: here we find $E_{4}>E_{3}$.

\section{Conclusion} \label{sec:V}
A one-parameter family of time-symmetric initial data for radial infall of a
particle into a Schwarzschild black hole was introduced in
perturbation theory.  The family is parameterized by $\alpha$, which measures the
gravitational-wave content of the initial surface.  The family contains a
conformally-flat initial three-geometry as a special member ($\alpha=1$).
Varying $\alpha$ allowed us to examine the influence of the initial data on
the gravitational waves emitted by the particle-black hole system.

We showed that for large initial separations, three stages can
be clearly identified in the radiative process: initial-data-produced
pulses, particle-produced acceleration radiation, and black-hole-produced
quasi-normal ringing.  For smaller
separations, the three stages become confused and interference takes place.

For large initial separations we
confirmed the general belief that a conformally-flat initial three-geometry
minimizes the radiated energy.  But we showed that for $r_{o}/2M<10$, the initial
configuration
that minimizes the energy is not the conformally-flat choice.
Instead, the configuration that minimizes the gravitational-wave content has
$\alpha<1$.

Most importantly, our numerical simulations show to what extent the gravitational
waveforms are influenced by the choice of initial data. As long as the particle falls toward the black hole
from a distance $r_{o}/2M > 10$, the part of the waveform that is associated with
acceleration radiation and quasi-normal ringing is insensitive to the choice of initial data.
In these cases, waveforms obtained with conformally-flat initial data are an accurate representation of
the true radiative process in the region of interest, because the unphysical radiation
coming from the initial data propagates to null-infinity before the physical radiation becomes important.
For $r_{o}/2M \lesssim 10$, however, the two contributions interfere, and uncertainties associated with the
choice of initial data hopelessly contaminate the waveforms.

\section*{Acknowledgement}
The authors gratefully acknowledge Thomas Baumgarte for useful correspondence
on numerical convergence.
This work was supported by the Natural Sciences and Engineering Research
Council of Canada.

\appendix*
\section{Convergence of the numerical code} \label{sec:convergence}
We now describe the convergence properties of our numerical algorithm.
The method described in Sec.~\ref{sec:nummet} is designed to be second-order
convergent, i.e. the numerical solution converges towards the exact solution quadratically.
This means that for a given choice of $r_{o}$, $\alpha$, and $l$, the field at a fixed spacetime
point is a function of $\Delta$ described by
\begin{eqnarray} \label{eqn:pd}
\psi_{N}(\Delta)=\psi_{A}+\Delta^2\rho,
\end{eqnarray}
where $\psi_N$ is the numerical field obtained with a stepsize $\Delta$ throughout the grid,
$\psi_{A}$ is the exact solution, and $\rho$ is an error term independent of $\Delta$.
Note that error terms of order $\mathcal{O}(\Delta^{3})$ are neglected in this equation.

The convergence of our numerical code can be tested by defining
\begin{eqnarray}\label{eqn:error}
\delta\psi(\Delta)=\psi_{N}(2\Delta)-\psi_{N}(\Delta)=3\Delta^2\rho,
\end{eqnarray}
and evaluating $\delta\psi$ for various values of $\Delta$.  This function
is proportional to the error made in discretizing Eq.~(\ref{eqn:wave}) and it satisfies
$\delta\psi(n\Delta)=n^2\delta\psi(\Delta)$.  This later property is important as we use it to
determine the convergence rate of our code.  We construct $\delta\psi$ from $\psi$ calculated
with $\Delta=0.005$, 0.01, 0.02, and 0.04 on the null line $v/2M=500$, i.e. at each grid point on this ingoing null line
we calculate $\delta\psi(0.005)$, $\delta\psi(0.01)$, and $\delta\psi(0.02)$.
In this way, we construct $\delta\psi(\Delta)$ as a function of retarded time $u$.  Note that for $u<-r^{*}_{o}$, we are testing the vacuum finite-difference
algorithm, whereas for $u>-r^{*}_{o}$ we are testing the convergence properties of the code involving the cells crossed
by the particle's world line.

\begin{figure}[t]
\vspace*{2.5in}
\special{hscale=31 vscale=31 hoffset=-13.0 voffset=190.0
         angle=-90.0 psfile=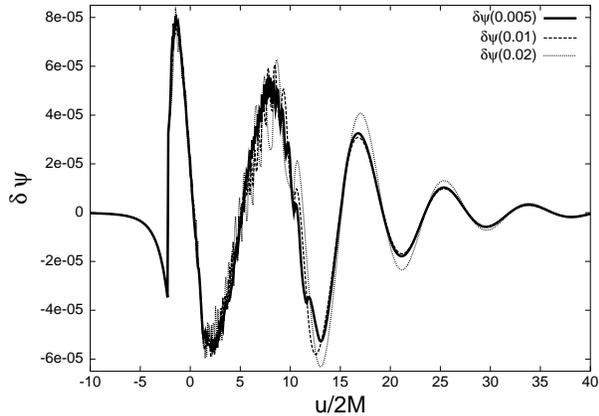}
\caption{Convergence properties of our numerical code. We display $\delta\psi(0.02)$, $4\delta\psi(0.01)$, and
$16\delta\psi(0.005)$, for $r_{o}/2M=2.1$, $\alpha=1$, and $l=2$.  The agreement between the curves indicates that
the numerical code is converging quadratically towards the exact solution.  This is characteristic of the convergence
in other cases.} \label{fig:conv}
\end{figure}

From the definition of $\delta\psi$, it is obvious
that they should satisfy $\delta\psi(0.02)=4\delta\psi(0.01)=16\delta\psi(0.005)$ at every grid point if Eq.~(\ref{eqn:pd}) holds and the convergence is quadratic.
In Fig.~\ref{fig:conv}, we display $\delta\psi(0.02)$, $4\delta\psi(0.01)$, and $16\delta\psi(0.005)$, for the case
 $r_{o}/2M=2.1$, $\alpha=1$, and $l=2$.  The fact that the three
curves are close together signals quadratic convergence, with $4\delta\psi(0.01)$ and $16\delta\psi(0.005)$ being
the closest.  This is expected since as $\Delta$ is decreased, the approximation made in Eq.~(\ref{eqn:pd})
becomes more accurate, and we get a better coincidence between the curves.
The convergence rate observed in this figure is typical of the convergence obtained
for other choices of $r_{o}$, $\alpha$, and $l$.

By varying the location of the event horizon, and the location
of future-null infinity, we have verified that our results are independent
of the actual location of the null boundaries.  We also verified that our results
for $\alpha=1$ are consistent with the results presented by Lousto and Price\cite{LP1}.

\end{document}